\documentclass[a4paper,11pt]{article}
\usepackage{a4wide,amsmath,amssymb,slashed,bbm}
\usepackage[english]{babel}

\usepackage{cite}

\usepackage{scalerel}
\newlength\bshft
\def\fakebold#1{\ThisStyle{\ooalign{$\SavedStyle#1$\cr%
  \kern-\bshft$\SavedStyle#1$\cr%
  \kern\bshft$\SavedStyle#1$}}}

\renewcommand{\u}{\underline}
\makeatletter

\@addtoreset{equation}{section} \makeatother

\begin{document}

\hfill{Imperial-TP-LW-2016-04}

\vspace{30pt}

\begin{center}
{\huge{\bf All symmetric space solutions of eleven-dimensional supergravity}}

\vspace{50pt}

Linus Wulff

\vspace{15pt}

{\it\small Blackett Laboratory, Imperial College, London SW7 2AZ, U.K.}\\

\vspace{100pt}

{\bf Abstract}
\end{center}
\noindent
We find all symmetric space solutions of eleven-dimensional supergravity completing an earlier classification by Figueroa-O'Farrill. They come in two types: AdS solutions and pp-wave solutions. We analyze the supersymmetry conditions and show that out of the 99 AdS geometries the only supersymmetric ones are the well known backgrounds arising as near-horizon limits of (intersecting) M-branes and preserving 32, 16 or 8 supersymmetries. The general form of the superisometry algebra for symmetric space backgrounds is also derived.

\pagebreak 
\tableofcontents

\setcounter{page}{1}


\section{Introduction}
Eleven-dimensional supergravity \cite{Cremmer:1978km} has a long history going back almost 40 years. Solutions to this theory are of great interest in string theory since eleven-dimensional supergravity is the low-energy limit of M-theory. In particular, solutions involving Anti-de Sitter space are of interest as they can provide examples of the AdS/CFT correspondence \cite{Maldacena:1997re} between M-theory in the AdS space and a dual boundary conformal field theory.\footnote{Note however that the vast majority of solutions we find are non-supersymmetric and recent arguments \cite{Ooguri:2016pdq,Freivogel:2016qwc} suggest that they should be unstable. A notable exception is $AdS_5\times\mathbbm{CP}^2\times T^2$ whose reduction to ten dimensions, although non-supersymmetric, is T-dual to $AdS_5\times S^5$ \cite{Duff:1998us}.} The most well understood example being $AdS_4\times S^7$ \cite{Aharony:2008ug}.

Here we will deal with a particularly simple class of solutions: symmetric spaces.\footnote{We assume that the four-form flux present in eleven-dimensional supergravity respects the symmetric space structure, i.e. it is given by a combination of invariant forms on the symmetric space.} These solutions are particularly easy to analyze as the differential equations of supergravity reduce to algebraic equations in this case. Besides their simplicity another reason to be interested in symmetric space solutions is that when reduced to ten dimensions they lead to integrable string theories, at least if the background preserves some amount of supersymmetry \cite{Wulff:2014kja,Wulff:2015mwa}. From this point of view it would be even more interesting to classify symmetric space solution of ten-dimensional supergravity but since this is a more difficult problem we will leave it for the future (see however \cite{FigueroaO'Farrill:2012rf}).

The problem of finding the symmetric space solutions of eleven-dimensional supergravity was addressed by Figueroa O'Farrill a couple of years ago in a very nice paper \cite{FigueroaO'Farrill:2011fj}. He showed that the solutions come in two families: pp-wave (or Cahen-Wallach) solutions and AdS solutions. He gave a list of all the AdS geometries that can occur and determined the fluxes completely for all cases except backgrounds of the form $AdS_2\times(S^2)^n\times T^{9-2n}$ for which the analysis was too involved. The supersymmetry of the backgrounds listed in \cite{FigueroaO'Farrill:2011fj} was later analyzed in \cite{Hustler:2015lca}.

Here we will revisit this problem and determine all symmetric space solutions of eleven-dimensional supergravity including the $AdS_2\times(S^2)^n\times T^{9-2n}$ ones. On the way we will correct some mistakes in the analysis of \cite{FigueroaO'Farrill:2011fj}. We will also analyze the supersymmetry of the AdS solutions completing the analysis of \cite{Hustler:2015lca}. The result is presented in tables \ref{tab:ads7}--\ref{tab:ads2}. Factors in the geometry with flux on them are underlined. The notation $S^2(H^2)$ means that either the compact space $S^2$ or the non-compact version $H^2$ can occur depending on parameters in the flux. We also give the number of supersymmetries preserved and the number of free parameters in the flux which we refer to as $F$ moduli.\footnote{Note that our analysis is purely classical and does not take into account for example quantization of brane charges. Our analysis also does not distinguish between compact and non-compact flat directions, i.e. $T^n$ vs $\mathbbm R^n$.} In total we find 99 AdS geometries
\begin{center}
\begin{itemize}
	\item[]3 $AdS_7$ geometries
	\item[]0 $AdS_6$ geometries
	\item[]9 $AdS_5$ geometries
	\item[]9 $AdS_4$ geometries
	\item[]25 $AdS_3$ geometries
	\item[]53 $AdS_2$ geometries
\end{itemize}
\end{center}
Comparing to the results of \cite{FigueroaO'Farrill:2011fj} we find that some solutions were missed there, namely solution 7 of table \ref{tab:ads5} and 22 of table \ref{tab:ads3} while solution 6 is more general and only the first branch of solution 60 of table \ref{tab:ads2} was found there. As already mentioned solutions of the form $AdS_2\times(S^2)^n\times T^{9-2n}$ for $n=1,2,3$ (40, 41, 43--45 of table \ref{tab:ads2}) were not analyzed in \cite{FigueroaO'Farrill:2011fj} due to the complexity of the flux. In addition \cite{FigueroaO'Farrill:2011fj} lists some spaces\footnote{They are: $AdS_2\times G_{\mathbbm C}(2,4)\times S^1$, $AdS_2\times S^3\times S^3\times T^3$, $AdS_2\times\mathbbm{CH}^2\times S^3\times T^2$, $AdS_2\times S^3\times H^2\times T^4$ and $AdS_2\times S^2\times S^2\times S^2(H^2)\times S^2(H^2)\times S^1$.} which we find not to solve the supergravity equations.

We also carry out a supersymmetry analysis for the complete list of symmetric space AdS backgrounds, tables \ref{tab:ads7}--\ref{tab:ads2}. Though most of the backgrounds were already analyzed in \cite{Hustler:2015lca} the present analysis has the advantage that it is not done by computer and is thus more transparent. The only supersymmetric solutions turn out to be the well known ones arising as near-horizon geometries of (intersecting) M-branes \cite{Gibbons:1993sv,Papadopoulos:1996uq,Tseytlin:1996bh,Gauntlett:1996pb,Tseytlin:1996hi,Cowdall:1998bu,Boonstra:1998yu}. These are listed in table \ref{tab:susy-solutions} together with their superisometry algebras and Killing spinor projection conditions. The general form of the superisometry algebra for symmetric space solutions\footnote{For a recent discussion of the superisometry algebra for general backgrounds see \cite{Figueroa-O'Farrill:2016wft}.} in terms of the geometry and flux is presented in (\ref{eq:alg}), whose form is formally the same as in the type II case \cite{Wulff:2015mwa}. This algebra reduces to those listed in table \ref{tab:susy-solutions} upon plugging in the corresponding curvature and flux. The superisometry algebra defines a supercoset space which is obtained from the full $(11|32)$-dimensional superspace by setting the Grassmann odd coordinates not associated to supersymmetries to zero as we show in appendix \ref{app:iso}. The supermembrane action restricted to this supercoset space is easily written down.

The outline of the paper is as follows. In the next section we analyze the algebraic problem of finding all symmetric space solutions of eleven-dimensional supergravity. Then in sec. \ref{sec:susy} we check the integrability condition for the Killing spinor equation to determine which of the AdS solutions are supersymmetric. Appendix \ref{app:sugra} gives the superspace constraints of eleven-dimensional supergravity and the derivation of the equations of motion from these. Appendix \ref{app:iso} gives the derivation of the superisometry algebra and supercoset subspace for a general symmetric space solution. Finally, in appendix \ref{app:ads2-s2}, we prove a useful result for classifying the most complicated $AdS_2$ solutions which were not previously analyzed.

\begin{table}[ht]
\begin{center}
\begin{tabular}{rlcc}
& Solution & Supersymmetries & $F$ moduli\\
\hline
1.&$AdS_7\times\u{S^4}$ & 32 & -\\
2.&$AdS_7\times\u{\mathbbm{CP}^2}$ & - & -\\
3.&$AdS_7\times\u{S^2\times S^2}$ & - & -
\end{tabular}
\caption{$AdS_7$ solutions. The underline indicates where the four-form flux is sitting.}
\label{tab:ads7}
\end{center}
\end{table}

\begin{table}[ht]
\begin{center}
\begin{tabular}{rlcc}
 & Solution & Supersymmetries & $F$ moduli\\
\hline
4.&$AdS_5\times\u{\mathbbm{CP}^3}$ & - & -\\
5.&$AdS_5\times\u{G_{\mathbbm R}^+(2,5)}$ & - & -\\
6.&$AdS_5\times\u{\mathbbm{CP}^2\times S^2(H^2)}$ & - & 1\\
7.&$AdS_5\times\u{\mathbbm{CP}^2\times T^2}$ & - & -\\
8.&$AdS_5\times\u{S^2\times S^2\times S^2(H^2)}$ & - & 2\\
9.&$AdS_5\times\u{S^2\times S^2\times T^2}$ & - & 1\\
10.&$AdS_5\times\u{S^4}\times H^2$ & - & -\\
%
\end{tabular}
\caption{$AdS_5$ solutions. The underline indicates where the four-form flux is sitting. The notation $S^2(H^2)$ means that this factor can be either the compact two-sphere or the corresponding non-compact space, the hyperbolic plane, depending on the value of the parameter(s) controlling the four-form flux.}
\label{tab:ads5}
\end{center}
\end{table}

\begin{table}[ht]
\begin{center}
\begin{tabular}{rlcc}
& Solution & Supersymmetries & $F$ moduli\\
\hline
11.&$\u{AdS_4}\times S^7$ & 32 & -\\
12.&$\u{AdS_4}\times S^5\times S^2$ & - & -\\
13.&$\u{AdS_4}\times SLAG_3\times S^2$ & - & -\\
14.&$\u{AdS_4}\times S^4\times S^3$ & - & -\\
15.&$\u{AdS_4}\times\mathbbm{CP}^2\times S^3$ & - & -\\
16.&$\u{AdS_4}\times S^3\times S^2\times S^2$ & - & -\\
17.&$AdS_4\times\u{S^4}\times H^3$ & - & -\\
18.&$AdS_4\times\u{\mathbbm{CP}}^2\times H^3$ & - & -\\
19.&$AdS_4\times\u{S^2\times S^2}\times H^3$ & - & -
\end{tabular}
\caption{$AdS_4$ solutions. The underline indicates where the four-form flux is sitting.}
\label{tab:ads4}
\end{center}
\end{table}

\begin{table}[ht]
\begin{center}
\begin{tabular}{rlcc}
& Solution & Supersymmetries & $F$ moduli\\
\hline
20.&$\u{AdS_3\times S^3\times S^3\times S^1}\times S^1$ & 16 & 1\\
21.&$\u{AdS_3\times S^3\times S^1}\times T^4$ & 16 & -\\
22.& $AdS_3\times\underline{\mathbbm{CP}^2\times\mathbbm{CP}^2(\mathbbm{CH}^2)}$ & - & 1\\
23.&$\left\{\begin{array}{l}AdS_3\times\u{\mathbbm{CH}^2\times S^2\times S^2}\\ AdS_3\times\u{\mathbbm{CP}^2\times S^2(H^2)\times S^2(H^2)}\end{array}\right.$ & - & 2\\
24.&$AdS_3\times\u{\mathbbm{CP}^2\times S^2(H^2)\times T^2}$ & - & 1\\
25.&$AdS_3\times\u{\mathbbm{CP}^2\times T^4}$ & - & -\\
26.&$AdS_3\times\u{S^2\times S^2\times S^2(H^2)\times S^2(H^2)}$ & - & 4\\
27.&$AdS_3\times\u{S^2\times S^2\times S^2(H^2)\times T^2}$ & - & 3\\
28.&$AdS_3\times\u{S^2\times S^2\times T^4}$ & -/-/-/8 & 2/2/2/1\\
29.&$AdS_3\times\u{S^2\times T^6}$ & 8 & -\\
30.&$AdS_3\times\u{S^4}\times H^4$ & - & -\\
31.&$AdS_3\times\u{\mathbbm{CP}}^2\times H^4$ & - & -\\
32.&$AdS_3\times\u{S^2\times S^2}\times H^4$ & - & -\\
33.&$AdS_3\times\u{S^4}\times\mathbbm{CH}^2$ & - & -\\
34.&$AdS_3\times\u{S^4}\times H^2\times H^2$ & - & -\\
%
35.&$AdS_3\times\u{\mathbbm{CP}^3}\times H^2$ & - & -\\
36.&$AdS_3\times\u{G_{\mathbbm R}^+(2,5)}\times H^2$ & - & -\\
\end{tabular}
\caption{$AdS_3$ solutions. The underline indicates where the four-form flux is sitting. For 28 there are four different branches of solutions, three non-supersymmetric and one supersymmetric, whose four-form flux involves respectively two and one free parameters.}
\label{tab:ads3}
\end{center}
\end{table}

\begin{table}[!ht]
\begin{center}
\begin{tabular}{rlcc}
& Solutions & Supersymmetries & $F$ moduli\\
\hline
37.&$\u{AdS_2\times\mathbbm{CP}^3\times T^2}\times S^1$ & - & -\\
38.&$\u{AdS_2\times G_{\mathbbm R}^+(2,5)\times T^2}\times S^1$ & - & -\\
39.&$\u{AdS_2\times\mathbbm{CP}^2\times S^2\times T^2}\times S^1$ & - & 1\\
40.&$\u{AdS_2\times S^2\times S^2\times S^2\times T^2}\times S^1$ & - & 2\\
41.&$\u{AdS_2\times S^2\times S^2\times S^2(H^2)}\times T^3$ & - & 3\\
42.&$\u{AdS_2\times\mathbbm{CP}^2\times T^2}\times T^3$ & - & 1\\
43.&$\u{AdS_2\times S^2\times S^2\times T^4}\times S^1$ & -/8/-/-/-/- & 3/1/3/2/3/2\\
44.&$\u{AdS_2\times S^2\times S^2\times T^2}\times T^3$ & - & 2\\
45.&$\u{AdS_2\times S^2\times T^4}\times T^3$ & -/-/8 & 1/1/-\\
46.&$AdS_2\times\u{SLAG_4}$ & - & -\\
47.&$\u{AdS_2\times H^2}\times S^7$ & - & -\\
48.&$\left\{\begin{array}{l}\u{AdS_2\times\mathbbm{CP}^2(\mathbbm{CH}^2)}\times S^5\\{}\u{AdS_2\times\mathbbm{CP}^2}\times H^5\end{array}\right.$ & - & 1\\
49.&$\left\{\begin{array}{l}\u{AdS_2\times\mathbbm{CP}^2(\mathbbm{CH}^2)}\times SLAG_3\\{}\u{AdS_2\times\mathbbm{CP}^2}\times SL(3,\mathbbm R)/SO(3)\end{array}\right.$ & - & 1\\
50.&$\left\{\begin{array}{l}\u{AdS_2\times S^2(H^2)\times S^2(H^2)}\times S^5\\{}\u{AdS_2\times S^2\times S^2}\times H^5\end{array}\right.$ & - & 2\\
51.&$\left\{\begin{array}{l}\u{AdS_2\times S^2(H^2)\times S^2(H^2)}\times SLAG_3\\{}\u{AdS_2\times S^2\times S^2}\times SL(3,\mathbbm R)/SO(3)\end{array}\right.$ & - & 2\\
52.&$\u{AdS_2\times S^2(H^2)\times T^2}\times S^5$ & - & 1\\
53.&$\u{AdS_2\times S^2(H^2)\times T^2}\times SLAG_3$ & - & 1\\
54.&$\u{AdS_2\times T^4}\times S^5$ & - & -\\
55.&$\u{AdS_2\times T^4}\times SLAG_3$ & - & -\\
56.&$\left\{\begin{array}{l}\u{AdS_2\times\mathbbm{CP}^2\times S^2(H^2)}\times S^3(H^3)\\{}\u{AdS_2\times\mathbbm{CH}^2\times S^2}\times S^3\end{array}\right.$ & - & 2\\
57.&$\left\{\begin{array}{l}\u{AdS_2\times S^2\times S^2(H^2)\times S^2(H^2)}\times S^3\\{}\u{AdS_2\times S^2\times S^2\times S^2(H^2)}\times H^3\end{array}\right.$ & - & 4\\
58.&$\u{AdS_2\times\mathbbm{CP}^2\times T^2}\times S^3(H^3)$ & - & 1\\
59.&$\left\{\begin{array}{l}\u{AdS_2\times S^2\times S^2(H^2)\times T^2}\times S^3\\{}\u{AdS_2\times S^2\times S^2\times T^2}\times H^3\end{array}\right.$ & - & 3\\
60.&$\u{AdS_2\times S^2\times T^4}\times S^3$ & -/-/-/-/8 & 2/2/2/2/1\\
61.&$\u{AdS_2\times T^6}\times S^3$ & 8 & -\\
%
62.&$\underline{AdS_2\times H^2}\times S^4\times S^3$ & - & -\\
63.&$AdS_2\times\underline{S^4}\times H^5$ & - & -\\
64.&$AdS_2\times\underline{S^4}\times SL(3,\mathbbm R)/SO(3)$ & - & -\\
65.&$AdS_2\times\underline{S^4}\times H^3\times H^2$ & - & -\\
66.&$AdS_2\times\underline{\mathbbm{CP}^3}\times H^3$ & - & -\\
67.&$AdS_2\times\underline{G_{\mathbbm R}^+(2,5)}\times H^3$ & - & -\\
\end{tabular}
\caption{$AdS_2$ solutions. The underline indicates where the four-form flux is sitting. When there are different branches of solutions they are separated by '/'.}
\label{tab:ads2}
\end{center}
\end{table}

\begin{table}[ht]
\begin{tabular}{lccc}
Solution & superisometry algebra & $\mathcal P$ ($\mathcal P\xi=\xi$)\\
\hline
$AdS_7\times S^4$ & $\mathfrak{osp}(2,6|4)$ & 1 \\
$AdS_4\times S^7$ & $\mathfrak{osp}(8|4)$ & 1 \\
$AdS_3\times S^3\times S^3\times T^2$ & $\mathfrak{d}(2,1;\alpha)\oplus\mathfrak{d}(2,1;\alpha)\oplus\mathbbm{R}^2$ & eq. (\ref{eq:proj-20}) \\
$AdS_3\times S^3\times T^5$ & $\mathfrak{psu}(1,1|2)\oplus\mathfrak{psu}(1,1|2)\oplus\mathbbm{R}^5$ & $\frac12(1-\Gamma^{012345})$ \\
$AdS_3\times S^2\times S^2\times T^4$ & $\mathfrak{d}(2,1;\alpha)\oplus\mathfrak{sl}(2,\mathbbm R)\oplus\mathbbm{R}^4$ & eq. (\ref{eq:proj-28}) \\
$AdS_3\times S^2\times T^6$ & $\mathfrak{psu}(1,1|2)\oplus\mathfrak{sl}(2,\mathbbm R)\oplus\mathbbm{R}^6$ & $\frac12(1+\Gamma^{3456})\frac12(1-\Gamma^{5678})$ \\
$AdS_2\times S^3\times S^2\times T^4$ & $\mathfrak{d}(2,1;\alpha)\oplus\mathfrak{su}(2)\oplus\mathbbm{R}^4$ & eq. (\ref{eq:proj-60}) \\
$AdS_2\times S^3\times T^6$ & $\mathfrak{psu}(1,1|2)\oplus\mathfrak{su}(2)\oplus\mathbbm{R}^6$ & $\frac12(1+\Gamma^{2345})\frac12(1-\Gamma^{4567})$ \\
$AdS_2\times S^2\times S^2\times T^5$ & $\mathfrak{d}(2,1;\alpha)\oplus\mathbbm{R}^5$ & eq. (\ref{eq:proj-43}) \\
$AdS_2\times S^2\times T^7$ & $\mathfrak{psu}(1,1|2)\oplus\mathbbm{R}^7$ & $\frac12(1-\Gamma^{012378})\frac12(1-\Gamma^{6789})$ \\
\end{tabular}
\caption{Supersymmetric AdS solutions of symmetric space type. The corresponding Killing spinor projection condition and superisometry algebras are listed.}
\label{tab:susy-solutions}
\end{table}

\section{Symmetric space solutions of eleven-dimensional supergravity}
The equations of motion of eleven-dimensional supergravity take the form\footnote{These are derived in the superspace formulation in appendix \ref{app:sugra}. In our conventions $R_{ab}=R_{ac}{}^c{}_b$, $i_aF=\frac16E^d\wedge E^c\wedge E^bF_{abcd}$ and the inner product of forms is defined as
$$
\langle\alpha,\beta\rangle=\frac{1}{n!}\alpha^{a_1\cdots a_n}\beta_{a_1\cdots a_n}\,,\qquad |\alpha|^2=\langle\alpha,\alpha\rangle\,.
$$
where $\alpha=\frac{1}{n!}E^{a_n}\wedge\cdots\wedge E^{a_1}\alpha_{a_1\cdots a_n}$ and similarly for $\beta$. The Hodge-dual of $F$ has components $(*F)_{fghijkl}=\frac{1}{4!}\varepsilon_{fghijklbcde}F^{bcde}$.
}
\begin{equation}
R_{ab}=\tfrac12\langle i_aF,i_bF\rangle-\tfrac16|F|^2\eta_{ab}\,,\qquad d*F=-\tfrac12F\wedge F\,.
\label{eq:Einstein}
\end{equation}
where $F=dC$, with $C$ the three-form potential of eleven-dimensional supergravity.

Specializing to symmetric space solutions $F$ must be constant, $d*F=0$, so that the corresponding equation of motion reduces to the algebraic equation $F\wedge F=0$. The Einstein equation also becomes algebraic by noting that for an irreducible symmetric space any symmetric bilinear form, e.g. the Ricci tensor, is proportional to the metric. The proportionality constant is then determined in terms of $F$ by the Einstein equation.

\begin{table}[ht]
\begin{center}
\begin{tabular}{r|l|l}
dim & Name & Invariant forms\\
\hline 2 & $S^2$ & 0,2\\
3 & $S^3$ & 0,3\\
4 & $\mathbbm{CP}^2$ & 0,2,4\\
4 & $S^4$ & 0,4\\
5 & $SLAG_3$ & 0,5\\
5 & $S^5$ & 0,5\\
6 & $\mathbbm{CP}^3$ & 0,2,4,6\\
6 & $G_{\mathbbm R}^+(2,5)$ & 0,2,4,6\\
6 & $S^6$ & 0,6\\
7 & $S^7$ & 0,7\\
8 & $\mathbbm{CP}^4$ & 0,2,4,6,8\\
8 & $G_{\mathbbm C}(2,4)$ & 0,2,4$^2$,6,8\\
8 & $S^8$ & 0,8\\
8 & $\mathbbm{HP}^2$ & 0,4,8\\
8 & $ASSOC$ & 0,4,8\\
8 & $SU(3)$ & 0,3,5,8\\
9 & $SLAG_4$ & 0,4,5,9\\
9 & $S^9$ & 0,9\\
10 & $\mathbbm{CP}^5$ & 0,2,4,6,8,10\\
10 & $G_{\mathbbm R}^+(2,7)$ & 0,2,4,6,8,10\\
10 & $S^{10}$ & 0,10\\
10 & $Sp(2)$ & 0,3,7,10
\end{tabular}
\caption{Irreducible Riemannian symmetric spaces up to dimension ten. For each entry there is a compact and a non-compact version. The name listed is that of the compact version. $G_{\mathbbm R}^+(k,n)$ is the Grassmannian of oriented $k$-planes in $\mathbbm{R}^n$, $G_{\mathbbm C}^+(k,n)$ the Grassmannian of complex $k$-planes in $\mathbbm{C}^n$, $ASSOC$ the Grassmannian of associative 3-planes in $\mathbbm{R}^7$ and $SLAG_n$ is the Grassmannian of special Lagrangian planes in $\mathbbm{C}^n$.}
\label{tab:riemannian}
\end{center}
\end{table}

A general symmetric space solution takes the form
\begin{equation}
\mathcal M_0\times\mathcal M_1\times\ldots\times\mathcal M_n
\end{equation}
where each $\mathcal M_i$ $i=1,\ldots,n$ is an irreducible Riemannian symmetric space and $\mathcal M_0$ is an (indecomposable) Lorentzian symmetric space. The Riemannian symmetric spaces were classified in terms of Lie algebras by Cartan and in table \ref{tab:riemannian} we list the irreducible ones up to dimension ten (the corresponding Lie algebras can be found in table 1 of \cite{FigueroaO'Farrill:2011fj}). The indecomposable Lorentzian symmetric spaces have been classified in \cite{Cahen-Wallach} and there are three classes: Anti-de Sitter space $AdS_n$, de Sitter space $dS_n$ and Cahen-Wallach (pp-wave) spaces\footnote{These spaces have metric
$$
ds^2=2dx^+dx^-+A_{ij}x^ix^j(dx^-)^2+dx^idx^i\,,
$$
with $A_{ij}$ a non-degenerate symmetric bilinear form which can be taken to be diagonal $A=\mathrm{diag}(a_1,\ldots,a_n)$. The only non-vanishing component of the Ricci tensor is $R_{--}=-\mathrm{tr}\,A$. The invariant forms are the constants together with $c_{i_1\cdots i_n} dx^-\wedge dx^{i_1}\wedge\cdots\wedge dx^{i_n}$ with $c_{i_1\cdots i_n}$ constant. See \cite{FigueroaO'Farrill:2011fj} for further details.} $CW_n(A)$ where $A$ is a symmetric bilinear form on the transverse $(n-2)$-dimensional subspace. It is easy to show that de Sitter solutions are not compatible with the supergravity equations, see \cite{FigueroaO'Farrill:2011fj} for a proof. It is also easy to show that for the Cahen-Wallach case the only supergravity solutions are of the form
\begin{equation}
CW_n(A)\times\mathbbm{R}^{11-n}\qquad F=dx^-\wedge\varphi\qquad\mathrm{tr}\,A=-\tfrac12|\varphi|^2\,,
\end{equation}
where $\varphi$ is a transverse constant coefficient three-form. The special case $n=2$ $F=0$ is eleven-dimensional Minkowski space. For the proof we refer again to \cite{FigueroaO'Farrill:2011fj}.

We are therefore left with the task of classifying symmetric space $AdS$ solutions of eleven-dimensional supergravity. This is considerably more involved and the end result is presented in tables \ref{tab:ads7}--\ref{tab:ads2}. Our calculations follow the same lines as \cite{FigueroaO'Farrill:2011fj} but will cover also the cases left out in that paper. We will also correct a few mistakes that appeared in that analysis. We will use a similar but not identical notation and we will organize the calculations in a different way. We will first analyze the backgrounds with $|F|^2=0$ and then the backgrounds with $|F|^2\neq0$. The latter case will be further subdivided in backgrounds without flux on AdS but flux on the rest of the space, backgrounds without flux on a single (indecomposable) Riemannian factor but flux on the rest of the space and remaining ones. As we will see in sec. \ref{sec:susy} the last class cannot give rise to supersymmetric backgrounds.

All solutions are given in equations with the geometry in boldface and the corresponding flux. The conditions on the parameters appearing in the flux are given just below, e.g. (\ref{eq:F-20}).

We use the following notation for forms
\begin{itemize}
	\item $\nu$ denotes the volume form on AdS with $|\nu|^2=-1$
	\item $\sigma$ denotes the volume form on spheres with $|\sigma|^2=1$
	\item $\omega$ denotes the K\"ahler form on $\mathbbm{CP}^n$, $G_{\mathbbm R}^+(2,5)$ or $T^6$ with $|\omega|^2=n$, for $T^6$ we take $\omega=\vartheta^{12}+\vartheta^{34}+\vartheta^{56}$
	\item $\vartheta^{12\cdots n}$ denotes the basic forms on flat directions, i.e. $\vartheta^{12\cdots n}=dx^1\wedge dx^2\wedge\ldots\wedge dx^n$
	\item $\omega_\pm$ denotes two choices of K\"ahler form on $T^4$, $\omega_\pm=\vartheta^{12}\pm\vartheta^{34}$
	\item $\Omega_\pm$ denotes the corresponding holomorphic two-form, $\Omega_\pm=(\vartheta^1+i\vartheta^2)\wedge(\vartheta^3\pm i\vartheta^4)$ with $\mathrm{Re}\,\Omega_\pm=\vartheta^{13}\mp\vartheta^{24}$ and $\mathrm{Im}\,\Omega_\pm=\vartheta^{23}\pm\vartheta^{14}$
\end{itemize}

Before we look at specific backgrounds let us show that for $AdS_n\times\mathcal M_{11-n}$ the space $\mathcal M_{11-n}$ must have positive scalar curvature and must therefore contain at least one irreducible Riemannian factor of positive curvature. If $n>4$ $F$ cannot have legs along AdS so that $|F|^2\geq0$. From the Einstein equation the Ricci scalar of AdS becomes $-\frac{n}{6}|F|^2$ while that of $\mathcal M_{11-n}$ becomes $\frac{n+1}{6}|F|^2$ which is indeed positive. If $n=4$ we have $F=f\nu+F'$ and the Ricci scalars become $-\frac23(2f^2+|F'|^2)<0$ and $\frac16(7f^2+5|F'|^2)>0$ respectively. For $n=3$ we have $F=\nu\wedge\rho+F'$ with $\rho\in\Omega^1(\mathcal M_{11-n})$. The Ricci scalars become $-\frac12(2|\rho|^2+|F'|^2)<0$ and $\frac16(5|\rho|^2+4|F'|^2)>0$ respectively. Finally for $n=2$ we have $F=\nu\wedge\rho+F'$ with $\rho\in\Omega^2(\mathcal M_{11-n})$. The Ricci scalars become $-\frac13(2|\rho|^2+|F'|^2)<0$ and $\frac12(|\rho|^2+|F'|^2)>0$ respectively completing the proof.

Another useful observation is that solutions with a $\mathbbm{CP}^2$ factor can always be treated as special cases of solutions with an $S^2\times S^2$ factor. This is because the solution is only sensitive to the invariant forms and the invariant forms of the former can be thought of as generated by the combination $\omega\sim\sigma_1+\sigma_2$ of the latter. Using this fact saves us having to analyze cases with $\mathbbm{CP}^2$ factors separately. Note that the same is true also for $\mathbbm{CP}^3$, $G_{\mathbbm R}^+(2,5)$ which can be treated as special cases of $S^2\times S^2\times S^2$ but in this case there are so few examples that this observation is not very useful other than as a consistency check.

\subsection{Solutions with $|F|^2=0$}\label{sec:F2-zero}
In this case the Einstein equation becomes
\begin{equation}
R_{ab}=\tfrac12\langle i_aF,i_bF\rangle\,,
\label{eq:Einstein-Fzero}
\end{equation}
so that directions without flux are Ricci-flat and hence flat. Therefore the solutions in this class take the form
\begin{equation}
\underline{AdS_n\times\mathcal M_k}\times T^{11-n-k}\,,
\end{equation}
where $\mathcal M_k$ is a Riemannian symmetric space. Directions with $F$-flux are underlined. Clearly we have $n=2,3,4$ since otherwise it is not possible to put flux on the $AdS$-directions without breaking isometries. Similarly $\mathcal M_k$ must consist of irreducible factors with invariant forms of degree $\leq3$ (for a factor with only invariant four-forms the condition $F\wedge F=0$ cannot be satisfied). Looking at the list in table \ref{tab:riemannian} the possible factors are $S^{2,3}$, $\mathbbm{CP}^{2,3,4}$, $G_{\mathbbm R}^+(2,5)$, $G_{\mathbbm C}(2,4)$ and $SU(3)$ or their non-compact versions.

\subsubsection*{$\fakebold{AdS_4}$ solutions}
This case is ruled out as follows. The flux is of the form $F=f\nu+F'$ where $\nu$ is the volume form on $AdS_4$ and $f$ is a non-vanishing constant. The condition $F\wedge F=0$ then implies that $F'=0$ but then $|F|^2\neq0$.

\subsubsection*{$\fakebold{AdS_3}$ solutions}
In this case there must be at least one flat direction with flux to be able to have flux on $AdS_3$. The flux is of the form $F=\nu\wedge\vartheta^1+F'$. The condition $F\wedge F=0$ implies that $F'=\rho\wedge\vartheta^1$ so that $F=(\nu+\rho)\wedge\vartheta^1$ for some non-vanishing three-form $\rho$. Note that the Einstein equation in the 1-direction forces $|F|^2=0$ so this analysis accounts for all solutions with flux on $AdS_3$. The only irreducible symmetric space of dimension less than eight with an invariant three-form is $S^3$. If there is no $S^3$ factor we must have $\rho=\sum_i\rho_{(i)}\wedge\vartheta^i$ for some two-forms $\rho_{(i)}$. This case is ruled out as follows. The Einstein equation (\ref{eq:Einstein-Fzero}) says that the component of the Ricci tensor along one of the flat directions $\vartheta^i$ is a sum of (strictly) positive terms which contradicts the fact that this direction is flat. This leaves only the possibilities $\rho=\sigma$ or $\rho=\sigma_1+\sigma_2$ where $\sigma$ denotes the volume form on $S^3(H^3)$. From the Einstein equation we find that the scalar curvature of the corresponding factor in the geometry is positive ruling out $H^3$. This leaves us with the solutions
\begin{equation}
\fakebold{\underline{AdS_3\times S^3\times S^3\times S^1}\times S^1}\,,\qquad\fakebold{\underline{AdS_3\times S^3\times S^1}\times T^4}\,,\qquad F=(f_1\nu+f_2\sigma_1+f_3\sigma_2)\wedge\vartheta^1\,,
\label{eq:F-20}
\end{equation}
where the Einstein equation implies $f_1^2=f_2^2+f_3^2$. The second solution corresponds to $f_3=0$ and arises as a limit of the first.

\subsubsection*{$\fakebold{AdS_2}$ solutions}
The flux is of the form $F=\nu\wedge\rho+F'$ where $\rho$ is some non-vanishing two-form. Let us start by considering the irreducible eight- or nine-dimensional factors in table \ref{tab:riemannian}. The only spaces with invariant two-forms are $\mathbbm{CP}^4$ and $G_{\mathbbm C}(2,4)$. For $\mathbbm{CP}^4$ taking $\rho=\omega$ with $\omega$ the K\"ahler form the condition $F\wedge F=0$ implies $\omega\wedge F'=0$ which in turn implies $F'=0$ but then $|F|^2\neq0$. For $G_{\mathbbm C}(2,4)$ there is one invariant two-form $\omega$ and two invariant four-forms $\Omega^{(i)}$ $i=1,2$. Since they satisfy $\omega\wedge(\Omega^{(1)}-\Omega^{(2)})=0$ (see appendix A.4 of \cite{FigueroaO'Farrill:2011fj}) we can take $\rho=\omega$ and $F'=\Omega^{(1)}-\Omega^{(2)}$ to satisfy $\rho\wedge F'=0$, however $F'\wedge F'$ is then non-zero and we do not get a solution.\footnote{\cite{FigueroaO'Farrill:2011fj} incorrectly lists this as a solution.}

For the six-dimensional factors $G_{\mathbbm R}^+(2,5)$ and $\mathbbm{CP}^3$ we have three extra directions which can form an $S^2\times S^1$ or a $T^3$ and the condition $F\wedge F=0$ means that the flux must take the form $F=f\nu\wedge\omega+\nu\wedge\rho_1+\omega\wedge\rho_2$ with $f\rho_2=0$ where $\rho_1=\rho_2=\sigma$ or $\rho_
1,\rho_2\in\Omega^2(T^3)$. If $\rho_2=0$ we get $|F|^2\neq0$ so we must have instead $f=0$. The $S^2$ case is ruled out since its curvature is proportional to $|F|^2$ which vanishes. We can take $\rho_1=f_1\vartheta^{12}$ and then the Einstein equation in the 3-direction forces $\rho_2\propto\rho_1$ and we get the solutions
\begin{equation}
\fakebold{\underline{AdS_2\times G_{\mathbbm R}^+(2,5)\times T^2}\times S^1}\,,\qquad
\fakebold{\underline{AdS_2\times\mathbbm{CP}^3\times T^2}\times S^1}\,,\qquad 
F=f(\sqrt3\,\nu+\omega)\wedge\vartheta^{12}\,.
\end{equation}

Next let us rule out geometries with $S^3$ factors. The highest dimension irreducible factor we could have is then $\mathbbm{CP}^2$ but then $F\wedge F=0$ cannot be satisfied. The most we can have are two $S^3$ factors. The remaining directions must then be flat and the flux must take the form $F=f_1\nu\wedge\vartheta^{12}+f_2\sigma_1\wedge\vartheta^1+f_3\sigma_2\wedge\vartheta^1$ to satisfy $F\wedge F=0$. But then the Einstein equation in the flat directions cannot be satisfied. With one $S^3$ factor we must have $F=f_1\nu\wedge\vartheta^{12}+f_2\sigma\wedge\vartheta^1+\rho\wedge\vartheta^1$ for some three-form $\rho$. However the difference of the Einstein equations in the 1 and 2-directions implies $f_2^2+|\rho'|^2=0$, where $\rho'$ is the part of $\rho$ with no legs along $\vartheta^2$, implying that $f_2=0$ leading to a contradiction. Note that we did not need to use the fact that $|F|^2=0$ to rule out these geometries so we have in effect ruled out all solutions of the form $AdS_2\times S^3\times\mathcal M_6$ with flux on $S^3$.

Finally we have to analyze backgrounds with only $S^2$ factors (backgrounds with $\mathbbm{CP}^2$ factors will be covered as special cases as already mentioned). These backgrounds turn out to be the most difficult to analyze especially $AdS_2\times S^2\times S^2\times S^2\times T^3$, $AdS_2\times S^2\times S^2\times T^5$ and $AdS_2\times S^2\times T^7$ which were for this reason not analyzed in \cite{FigueroaO'Farrill:2011fj}. In appendix \ref{app:ads2-s2} we show that for these cases there is always a flat direction without flux and hence $|F|^2=0$, therefore the analysis in this section covers all such solutions.

We rule out solutions with four $S^2$ factors as follows.\footnote{Again due to a mistake \cite{FigueroaO'Farrill:2011fj} lists such solutions.} The flux would take the form
\begin{equation}
F=f_i\nu\wedge\sigma_i+\tfrac12f_{ij}\sigma_i\wedge\sigma_j\,,\qquad i,j=1,\ldots,4\,.
\end{equation}
We can assume, without loss of generality, that $f_1\neq0$. The condition $F\wedge F=0$ then implies the relations
\begin{align}
&f_{23}=-\frac{f_2f_{13}+f_3f_{12}}{f_1}\,,\qquad
f_{24}=-\frac{f_2f_{14}+f_4f_{12}}{f_1}\,,\qquad
f_{34}=-\frac{f_3f_{14}+f_4f_{13}}{f_1}\,,\nonumber\\
&f_2f_3f_{14}+f_2f_4f_{13}+f_3f_4f_{12}=0\,,\qquad
f_3f_{12}f_{14}+f_4f_{12}f_{13}+f_2f_{13}f_{14}=0\,.
\end{align}
Assume that $f_2=0$. Since all non-flat directions must have flux on them we find that $f_3=f_4=0$ and $f_{23}=f_{24}=f_{34}=0$ but then the curvature of the first $S^2$ is proportional to $|F|^2$ which vanishes giving a contradiction. Therefore $f_2,f_3,f_4$ must be non-zero and we find
\begin{align}
f_{14}=f_4\left(\frac{f_{12}}{f_2}+\frac{f_{13}}{f_3}\right)\,,\qquad
f_2f_3f_{12}f_{13}+f_3^2f_{12}^2+f_2^2f_{13}^2=0\,.
\end{align}
The only (real) solution to the last equation is $f_{12}=f_{13}=0$ which, together with $|F|^2=0$ implies that $F=0$ giving a contradiction. Since the analysis for backgrounds involving $\mathbbm{CP}^2\times\mathbbm{CP}^2$ and $\mathbbm{CP}^2\times S^2\times S^2$ are special cases of the above analysis (with $\omega\sim\sigma_1+\sigma_2$) these are also ruled out.

The same analysis, with $\sigma_1$ or $\sigma_4$ replaced by $\vartheta^{12}$, applies to the case with three $S^2$ factors (it is not hard to show that this is the most general form of $F$). In the first case it is now consistent to take $f_2=0$ giving the solution (where we renamed the $f$'s)
\begin{equation}
\fakebold{\underline{AdS_2\times S^2\times S^2\times S^2\times T^2}\times S^1}\,,\qquad
F=(f_1\nu+f_2\sigma_1+f_3\sigma_2+f_4\sigma_3)\wedge\vartheta^{12}\,,
\label{eq:F-40}
\end{equation}
where the condition $|F|^2=0$ implies that $f_1^2=f_2^2+f_3^2+f_4^2$. In the second case we find a solution with $f_4=f_{i4}=0$ leading to (renaming the $f$'s)
\begin{align}
&\fakebold{\underline{AdS_2\times S^2\times S^2\times S^2(H^2)}\times T^3}\,,
\nonumber\\
&F=f_1\nu\wedge(\sigma_1+f_2\sigma_2+f_3\sigma_3)
+f_4\sigma_1\wedge\sigma_2
+f_5\sigma_1\wedge\sigma_3
-(f_3f_4+f_2f_5)\sigma_2\wedge\sigma_3
\,.
\label{eq:F-41}
\end{align}
The condition $|F|^2=0$ implies the relation $f_1^2(1+f_2^2+f_3^2)=f_4^2+f_5^2+(f_2f_5+f_3f_4)^2$. The curvatures of the three $S^2(H^2)$'s are proportional to $f_4^2+f_5^2-f_1^2$, $f_1^2(1+f_3^2)-f_5^2$ and $f_1^2(1+f_2^2)-f_4^2$ and the sum of any two of these is positive so that at most one can be negative.

By taking $f_2=f_3$ in the first of the above solutions (\ref{eq:F-40}) we find the most general solution involving $\mathbbm{CP}^2\times S^2$ namely
\begin{equation}
\fakebold{\underline{AdS_2\times\mathbbm{CP}^2\times S^2\times T^2}\times S^1}\,,\qquad
F=(f_1\nu+f_2\omega+f_3\sigma)\wedge\vartheta^{12}\,,
\end{equation}
with $f_1^2=2f_2^2+f_3^2$ to satisfy $|F|^2=0$. At first sight it seems like we would get a different solution by taking $f_2=f_3$ and $f_4=f_5$ in the second solution above (\ref{eq:F-41}), but in that case one finds from the Einstein equations that the curvature of the $S^2$ vanishes giving a contradiction.

Next we analyze backgrounds of the form $AdS_2\times S^2\times S^2\times T^5$, which turns out to be the most difficult ones. Without loss of generality the flux takes the form
\begin{equation}
F=
f_1\nu\wedge\sigma_1
+f_2\nu\wedge\sigma_2
+\nu\wedge(f_3\vartheta^{12}+f_4\vartheta^{34})
+f_5\sigma_1\wedge\sigma_2
+\sigma_1\wedge\alpha
+\sigma_2\wedge\beta
+\gamma\,,
\end{equation}
with $\alpha,\,\beta\in\Omega^2(T^5)$ and $\gamma\in\Omega^4(T^5)$. The Einstein equation in the $5$-direction then implies that there cannot be any flux on this direction so that $\alpha$, $\beta$ and $\gamma$ only have legs on $T^4$ and we can take $\gamma=f_6\vartheta^{1234}$. The trace of the Einstein equations in the torus directions and the condition $|F|^2=0$ imply, respectively,
\begin{equation}
2f_3^2+2f_4^2=2|\alpha|^2+2|\beta|^2+4f_6^2\,,\qquad f_1^2+f_2^2+f_3^2+f_4^2=f_5^2+|\alpha|^2+|\beta|^2+f_6^2\,,
\end{equation}
which gives $f_5^2=f_1^2+f_2^2+f_6^2$. Let us write $\alpha=\frac12f_{ij}\vartheta^{ij}$ and $\beta=\frac12g_{ij}\vartheta^{ij}$ and note that the freedom of rotation in the $(12)$ and $(34)$ planes allows us to set $f_{14}=0$. The condition $F\wedge F=0$ then implies the relations
\begin{align}
&f_1g_{12}+f_2f_{12}+f_3f_5=0\,,\qquad
f_1g_{13}+f_2f_{13}=0\,,\qquad
f_1g_{14}=0\,,\qquad
f_1g_{23}+f_2f_{23}=0\,,
\nonumber\\
&f_1g_{24}+f_2f_{24}=0\,,\qquad
f_1g_{34}+f_2f_{34}+f_4f_5=0\,,\qquad
f_1f_6+f_3f_{34}+f_4f_{12}=0\,,
\nonumber\\
&f_2f_6+f_3g_{34}+f_4g_{12}=0\,,\qquad
f_5f_6+f_{12}g_{34}-f_{13}g_{24}+f_{23}g_{14}-f_{24}g_{13}+f_{34}g_{12}=0\,,
\label{eq:F-cond1-43}
\end{align}
and the diagonal components of the Einstein equation in the flat directions imply
\begin{align}
&f_3^2=f_{12}^2+f_{13}^2+g_{12}^2+g_{13}^2+g_{14}^2+f_6^2\,,\qquad
f_3^2-f_4^2=f_{12}^2-f_{34}^2+g_{12}^2-g_{34}^2\,,
\nonumber\\
&f_{24}^2+g_{24}^2=f_{13}^2+g_{13}^2\,,\qquad
f_{23}^2+g_{23}^2=g_{14}^2\,,
\label{eq:Einstein-cond1-43}
\end{align}
while the off-diagonal components imply
\begin{equation}
f_{ij}f_{jk}+g_{ij}g_{jk}=0\qquad i\neq k\,.
\label{eq:Einstein-cond1-43-off-diag}
\end{equation}

Let us first assume that $f_1=f_2=0$. Since $f_3$ and $f_4$ cannot both vanish the conditions $f_3f_5=f_4f_5=0$ imply $f_5=0$ which means that also $f_6=0$. We must then have $f_{3,4}\neq0$ and the $F\wedge F=0$ conditions (\ref{eq:F-cond1-43}) imply
\begin{equation}
f_{34}=-\frac{f_4f_{12}}{f_3}\,,\qquad
g_{34}=-\frac{f_4g_{12}}{f_3}\,,\qquad
\frac{2f_4f_{12}g_{12}}{f_3}+f_{13}g_{24}-f_{23}g_{14}+f_{24}g_{13}=0\,.
\end{equation}
The torus Einstein equations (\ref{eq:Einstein-cond1-43}) then imply, among others, the condition
\begin{equation}
(f_{12}^2+g_{12}^2-f_3^2)(f_3^2-f_4^2)=0\,.
\end{equation}
If the first factor vanishes we find from the remaining equations $f_{13}=f_{23}=f_{24}=g_{13}=g_{14}=g_{23}=g_{24}=0$ and $f_{12}g_{12}=0$ which implies that there is no flux on one $S^2$ giving a contradiction. Therefore the second factor must vanish so that $f_4=f_3$ (up to an inconsequential sign) and we get $f_{34}=-f_{12}$ and $g_{34}=-g_{12}$ together with the conditions
\begin{align}
&f_3^2=f_{12}^2+f_{13}^2+g_{12}^2+g_{13}^2+g_{14}^2\,,\qquad
f_{24}^2+g_{24}^2=f_{13}^2+g_{13}^2\,,\qquad
\nonumber\\
&
g_{14}^2=f_{23}^2+g_{23}^2\,,\qquad
2f_{12}g_{12}+f_{13}g_{24}-f_{23}g_{14}+f_{24}g_{13}=0\,,
\nonumber\\
&f_{23}(f_{13}-f_{24})+(g_{13}-g_{24})(g_{23}-g_{14})=0\,,\qquad
f_{12}f_{23}+g_{12}(g_{14}+g_{23})=0\,,\qquad
\nonumber\\
&f_{12}(f_{13}-f_{24})+g_{12}(g_{13}-g_{24})=0\,,\qquad
f_{23}f_{24}+g_{13}g_{14}+g_{23}g_{24}=0\,.
\end{align}
With some work one can show that there is no solution unless $f_{23}=0$. In that case there are 6 solutions (recall that $f_{14}=0$)
\begin{align}
(i)
&\qquad
g_{12}=g_{34}=0\,,\qquad
f_{24}=f_{13}\,,\qquad
g_{23}=g_{14}\,,\qquad
g_{24}=-g_{13}\,,
\\
(ii)
&\qquad
f_{12}=f_{34}=g_{12}=g_{34}=g_{13}=g_{24}=0\,,\qquad
f_{24}=-f_{13}\,,\qquad
g_{23}=g_{14}\,,
\\
(iii)
&\qquad
f_{12}=f_{34}=0\,,\qquad
f_{24}=-f_{13}\,,\qquad
g_{23}=-g_{14}\,,\qquad
g_{24}=g_{13}\,,
\\
(iv)
&\qquad
f_{12}=f_{34}=g_{13}=g_{24}=0\,,\qquad
f_{24}=f_{13}\,,\qquad
g_{23}=-g_{14}\,,
\\
(v)
&\qquad
f_{24}=f_{13}\,,\qquad
g_{12}=-f_{13}g_{13}/f_{12}\,,\qquad
g_{23}=-g_{14}\,,\qquad
g_{24}=g_{13}\,,
\\
(vi)
&\qquad
f_{12}=f_{34}=g_{14}=g_{23}=0\,,\qquad
f_{24}=f_{13}\,,\qquad
g_{24}=-g_{13}\,,
\end{align}
with $f_3^2=f_{12}^2+f_{13}^2+g_{12}^2+g_{13}^2+g_{14}^2$. This leads to the solutions
\begin{align}
\fakebold{\underline{AdS_2\times S^2\times S^2\times T^4}\times S^1}\,,\qquad
F=&f_3\nu\wedge\omega_++\sigma_1\wedge(f_{12}\omega_-+f_{13}\mathrm{Re}\,\Omega_-)
\nonumber\\
&{}
+\sigma_2\wedge(g_{13}\mathrm{Re}\,\Omega_++g_{14}\mathrm{Im}\,\Omega_+)\,,
\label{eq:F-43-1}
\end{align}
with $f_3^2=f_{12}^2+f_{13}^2+g_{13}^2+g_{14}^2$,
\begin{align}
\fakebold{\underline{AdS_2\times S^2\times S^2\times T^4}\times S^1}\,,\qquad
F=&f_3\nu\wedge\omega_++f_{13}\sigma_1\wedge\mathrm{Re}\,\Omega_++g_{14}\sigma_2\wedge\mathrm{Im}\,\Omega_+\,,
\label{eq:F-43-2}
\end{align}
with $f_3^2=f_{13}^2+g_{14}^2$,
\begin{align}
\fakebold{\underline{AdS_2\times S^2\times S^2\times T^4}\times S^1}\,,\qquad 
F=&f_3\nu\wedge\omega_++f_{13}\sigma_1\wedge\mathrm{Re}\,\Omega_+
\nonumber\\
&{}
+\sigma_2\wedge\big(g_{12}\omega_-+g_{13}\mathrm{Re}\,\Omega_--g_{14}\mathrm{Im}\,\Omega_-\big)\,,
\label{eq:F-43-3}
\end{align}
with $f_3^2=f_{13}^2+g_{12}^2+g_{13}^2+g_{14}^2$,
\begin{align}
\fakebold{\underline{AdS_2\times S^2\times S^2\times T^4}\times S^1}\,,\qquad
F=f_3\nu\wedge\omega_++f_{13}\sigma_1\wedge\mathrm{Re}\,\Omega_-+\sigma_2\wedge(g_{12}\omega_--g_{14}\mathrm{Im}\,\Omega_-)\,,
\label{eq:F-43-4}
\end{align}
with $f_3^2=f_{13}^2+g_{12}^2+g_{14}^2$, (redefining the coefficients slightly)
\begin{align}
\fakebold{\underline{AdS_2\times S^2\times S^2\times T^4}\times S^1}\,,\qquad
F=&f_3\nu\wedge\omega_++f_{12}\sigma_1\wedge(\omega_-+f_{13}\mathrm{Re}\,\Omega_-)
\nonumber\\
&{}
+\sigma_2\wedge(f_{13}g_{13}\omega_--g_{13}\mathrm{Re}\,\Omega_--g_{14}\mathrm{Im}\,\Omega_-)\,,
\label{eq:F-43-5}
\end{align}
with $f_3^2=(f_{12}^2+g_{13}^2)(1+f_{13}^2)+g_{14}^2$ and
\begin{align}
\fakebold{\underline{AdS_2\times S^2\times S^2\times T^4}\times S^1}\,,\qquad
F=f_3\nu\wedge\omega_+
+f_{13}\sigma_1\wedge\mathrm{Re}\,\Omega_-
+\sigma_2\wedge(g_{12}\omega_-+g_{13}\mathrm{Re}\,\Omega_+)\,,
\label{eq:F-43-6}
\end{align}
with $f_3^2=f_{13}^2+g_{12}^2+g_{13}^2$.

It remains to analyze the case $f_1\neq0$ say. We find from (\ref{eq:F-cond1-43}), (\ref{eq:Einstein-cond1-43}) and (\ref{eq:Einstein-cond1-43-off-diag})
\begin{align}
&g_{12}=-\frac{f_3f_5+f_2f_{12}}{f_1}\,,\qquad
g_{13}=-\frac{f_2f_{13}}{f_1}\,,\qquad
g_{14}=g_{23}=f_{23}=0\,,\qquad
g_{24}=-\frac{f_2f_{24}}{f_1}\,,
\nonumber\\
&g_{34}=-\frac{f_4f_5+f_2f_{34}}{f_1}\,,\qquad
f_{24}=\pm f_{13}\,,\qquad
f_1f_6+f_3f_{34}+f_4f_{12}=0\,,\qquad
f_1f_2f_6-f_3f_4f_5=0\,,
\nonumber\\
&f_1f_5f_6-f_2f_{12}f_{34}\pm f_2f_{13}^2=0\,,
\end{align}
together with
\begin{align}
&f_3^2=\left(1+\frac{f_2^2}{f_1^2}\right)(f_{12}^2+f_{13}^2)
+\frac{2f_2f_3f_5f_{12}}{f_1^2}
+\frac{f_3^2f_5^2}{f_1^2}
+f_6^2\,,\qquad
f_5^2=f_1^2+f_2^2+f_6^2\,,
\nonumber\\
&f_3^2-f_4^2=f_{12}^2-f_{34}^2+\frac{(f_3f_5+f_2f_{12})^2}{f_1^2}-\frac{(f_4f_5+f_2f_{34})^2}{f_1^2}\,,
\nonumber\\
&f_{13}[(f_1^2+f_2^2)(f_{12}\pm f_{34})+f_2f_5(f_3\pm f_4)]=0\,.
\end{align}
Assuming $f_2\neq0$ we get $f_6=f_3f_4f_5/f_1f_2$ and $f_5^2=(f_1^2+f_2^2)f_1^2f_2^2/(f_1^2f_2^2-f_3^2f_4^2)$ and defining $f_{ij}=f_5f_{ij}'/\sqrt{f_1^2+f_2^2}$, $f_3=f_3'\sqrt{f_1^2+f_2^2}$ and $f_4=f_4'\sqrt{f_1^2+f_2^2}$ the first equation above can be written as the vanishing of a sum of squares
\begin{equation}
f_2^2(f'_{12}+f_2f'_3)^2+f_2^2f_{13}'^2+(1+f_3'^2)(f_1^2+f_2^2)^2f_3'^2f_4'^2=0\,,
\end{equation}
which forces
\begin{equation}
f_3f_4=f_{13}=0\,,\qquad f_{12}=-\frac{f_2f_3f_5}{f_1^2+f_2^2}\,.
\end{equation}
The remaining equations then give (we take $f_4=0$ without loss of generality)
\begin{equation}
f_{12}=-\frac{f_2f_3}{f_5}\,,\qquad
g_{12}=-\frac{f_1f_3}{f_5}\,,\qquad
f_5^2=f_1^2+f_2^2\,,
\end{equation}
and the rest vanishing and we get the solution
\begin{equation}
\fakebold{\underline{AdS_2\times S^2\times S^2\times T^2}\times T^3}\,,\qquad
F=
\nu\wedge(f_1\sigma_1+f_2\sigma_2+f_3\vartheta^{12})
+f_5\sigma_1\wedge\sigma_2
-\frac{f_3}{f_5}(f_2\sigma_1+f_1\sigma_2)\wedge\vartheta^{12}\,,
\label{eq:F-44}
\end{equation}
with $f_5=\sqrt{f_1^2+f_2^2}$. The curvatures of the spheres are proportional to $f_2^2(1+f_3^2/f_5^2)$ and $f_1^2(1+f_3^2/f_5^2)$ respectively. This solution can be obtained from that in (\ref{eq:F-41}) by setting the curvature of one of the spheres to zero. By analyzing separately the case $f_2=0$ we find that it is a special case of the above solution, which is however ruled out as the curvature of one of the spheres vanishes. 

By considering special cases of the analysis above we obtain the following solution with a $\mathbbm{CP}^2$ factor
\begin{equation}
\fakebold{\underline{AdS_2\times\mathbbm{CP}^2\times T^2}\times T^3}\,,\qquad
F=\nu\wedge(f_1\omega+f_2\vartheta^{12})+\tfrac{1}{\sqrt2}\omega\wedge(f_1\omega-f_2\vartheta^{12})\,.
\label{eq:F-42}
\end{equation}

Finally, for $AdS_2\times S^2\times T^7$ the flux takes the form
\begin{equation}
F=f_1\nu\wedge\sigma+\nu\wedge(f_2\vartheta^{12}+f_3\vartheta^{34}+f_4\vartheta^{56})+\sigma\wedge\alpha+\beta\,,
\end{equation}
with $\alpha\in\Omega^2(T^7)$ and $\beta\in\Omega^4(T^7)$. The Einstein equation in the seventh torus direction implies that $\alpha$ and $\beta$ only have legs on $T^6$, i.e. there is no flux on the seventh torus direction. The trace of the Einstein equation in the torus directions and the fact that $|F|^2=0$ imply, respectively,
\begin{equation}
2f_2^2+2f_3^2+2f_4^2=2|\alpha|^2+4|\beta|^2\,,\qquad f_1^2+f_2^2+f_3^2+f_4^2=|\alpha|^2+|\beta|^2\,,
\end{equation}
which implies $f_1=0$ and $\beta=0$. Without loss of generality we can take $f_2\neq0$. The condition $F\wedge F=0$ then implies, setting $\alpha=\frac12f_{ij}\vartheta^{ij}$ with $f_{14}=0$, that
\begin{align}
&f_{35}=f_{36}=f_{45}=f_{46}=0\,,\qquad
f_{34}=-\frac{f_3f_{12}}{f_2}\,,\qquad
f_{56}=-\frac{f_4f_{12}}{f_2}\,,\qquad
f_3f_4f_{12}=0\,,\nonumber\\
&f_3f_{15}=f_3f_{16}=f_3f_{25}=f_3f_{26}=0\,,\qquad
f_4f_{13}=f_4f_{23}=f_4f_{24}=0\,.
\end{align}
If $f_3$ and $f_4$ are both non-vanishing we find that $\alpha=0$ but then the Einstein equation in the torus directions cannot be satisfied. Therefore, without loss of generality, we take $f_4=0$ and then the Einstein equation in the torus directions force $f_{i5}=f_{i6}=0$ as well as
\begin{equation}
f_2^2=f_{12}^2+f_{13}^2=f_{12}^2+f_{23}^2+f_{24}^2\,,\qquad
f_3^2=f_{13}^2+f_{23}^2+f_{34}^2=f_{24}^2+f_{34}^2\,,
\end{equation}
and
\begin{equation}
f_{ij}f_{jk}=0\qquad i\neq k\,,
\end{equation}
which implies
\begin{equation}
f_{24}=\pm f_{13}\,,\qquad
f_{23}=0\,,\qquad
f_2^2=f_{12}^2+f_{13}^2\,,\qquad
(f_3^2-f_2^2)f_{13}^2=0\,,\qquad
f_{13}(f_{12}\pm f_{34})=0\,.
\end{equation}
We get three solutions
\begin{equation}
\fakebold{\underline{AdS_2\times S^2\times T^4}\times T^3}\,,\qquad
F=\nu\wedge(f_2(\omega_++\omega_-)+f_3(\omega_+-\omega_-))+\sigma\wedge(f_2(\omega_++\omega_-)-f_3(\omega_+-\omega_-))\,,
\label{eq:F-45-1}
\end{equation}
which can be obtained by setting $f_2=0$ in the solution (\ref{eq:F-44}),
\begin{align}
\fakebold{\underline{AdS_2\times S^2\times T^4}\times T^3}\,,\qquad
F=f_2\nu\wedge\omega_++f_{12}\sigma\wedge\omega_-+f_{13}\sigma\wedge\mathrm{Re}\,\Omega_-\,,
\label{eq:F-45-2}
\end{align}
with $f_2^2=f_{12}^2+f_{13}^2$ and
\begin{align}
\fakebold{\underline{AdS_2\times S^2\times T^4}\times T^3}\,,\qquad
F=f_2\nu\wedge\omega_++f_2\sigma\wedge\mathrm{Re}\,\Omega_+\,.
\label{eq:F-45-3}
\end{align}
The last two solutions can be obtained by setting the flux on one $S^2$ to zero in the solutions (\ref{eq:F-43-1})--(\ref{eq:F-43-3}).


\subsection{Solutions with $|F|^2\neq0$: only AdS without flux}
As already mentioned it will be convenient to split the solutions with $|F|^2\neq0$ into three classes: (i) those with flux on all except the AdS directions, (ii) those with flux on all directions except for a single indecomposable Riemannian factor and (iii) the rest. Here we analyze the first class.

Note that since $|F|^2\neq0$ the Einstein equation, (\ref{eq:Einstein}), implies that all flat directions must have flux on them.

\subsubsection*{$\fakebold{AdS_7}$ solutions}
The possible geometries are
\begin{equation}
\fakebold{AdS_7\times\underline{S^4}}\,,\qquad
\fakebold{AdS_7\times\underline{\mathbbm{CP}^2}}\,,\qquad
\fakebold{AdS_7\times\underline{S^2\times S^2}}\,.
\end{equation}
Examples with flat directions, such as $S^3\times S^1$, are ruled out by the Einstein equation.

\subsubsection*{$\fakebold{AdS_6}$ solutions}
The only possibility for putting the $F$ flux on the five remaining directions is if they are of the form $S^3\times T^2$ or $S^2\times T^3$. However, by a rotation one can get rid of the flux on one of the flat directions giving a contradiction.

\subsubsection*{$\fakebold{AdS_5}$ solutions}
We clearly have the possibility of a six-dimensional irreducible factor
\begin{equation}
\fakebold{AdS_5\times\underline{\mathbbm{CP}^3}}\,,\qquad
\fakebold{AdS_5\times\underline{G_{\mathbbm R}^+(2,5)}}\,,\qquad F=f\omega\wedge\omega\,.
\end{equation}
We cannot have an $S^4$ factor since we can not put flux on the remaining two directions. The case of $\mathbbm{CP}^2$ will be treated as a special case of the analysis for $S^2\times S^2$. Backgrounds involving $S^3$ are easily seen to be ruled out since it is not possible to put flux on all except the $AdS_5$ directions. 

This leaves geometries with $S^2$ factors and flat directions and we find the possibilities
\begin{equation}
\fakebold{AdS_5\times\underline{S^2\times S^2\times S^2(H^2)}}\,,\qquad
\fakebold{AdS_5\times\underline{S^2\times S^2\times T^2}}\,,\qquad 
F=f_1\sigma_1\wedge\sigma_2+f_2\sigma_1\wedge\rho+f_3\sigma_2\wedge\rho\,,
\label{eq:F-8}
\end{equation}
where $\rho$ is either $\sigma_3$ or $\vartheta^{12}$. The curvatures of the two-dimensional factors are proportional to $2f_1^2+2f_2^2-f_3^3$, $2f_1^2-f_2^2+2f_3^3$ and $-f_1^2+2f_2^2+2f_3^3$ and the sum of any two is positive. The $T^2$ case has $f_1^2=2f_2^2+2f_3^2$. The case $S^2\times T^4$ is clearly ruled out.

A special case of the above analysis gives the solutions involving a $\mathbbm{CP}^2$ factor\footnote{Only the $f_2=0$ case appears in \cite{FigueroaO'Farrill:2011fj}.}
\begin{equation}
\fakebold{AdS_5\times\underline{\mathbbm{CP}^2\times S^2(H^2)}}\,,\qquad
\fakebold{AdS_5\times\underline{\mathbbm{CP}^2\times T^2}}\,,\qquad F=\tfrac12f_1\omega\wedge\omega+f_2\omega\wedge\rho\,,
\end{equation}
with $\rho$ either $\sigma$ or $\vartheta^{12}$. The $\mathbbm{CP}^2$ curvature is proportional to $2f_1^2+f_2^2$ while that of the two-dimensional factor is proportional to $4f_2^2-f_1^2$ leading to $S^2$ when this is positive, $H^2$ when it's negative and $T^2$ when $f_1^2=4f_2^2$.

\subsubsection*{$\fakebold{AdS_4}$ solutions}
The only irreducible seven-dimensional factor is $S^7$ which is clearly ruled out. Similarly any six-dimensional factor times $S^1$ is ruled out since there is no way to put flux on all directions. The same is true for five-dimensional factors as well as $S^4$. For the same reason also $\mathbbm{CP}^2$ is ruled out. For spaces involving $S^3$ factors such as $S^3\times S^3\times S^1$ with flux $F=(f_1\sigma_1+f_2\sigma_2)\wedge\vartheta^1$ it is not possible to satisfy the Einstein equation in the flat direction. The same is true for $S^3\times S^2\times T^2$ and $S^3\times T^4$. This leaves backgrounds with only $S^2$ factors and flat directions. With three $S^2$ factors it is not possible to put flux on the remaining flat direction. For $S^2\times S^2\times T^3$ we have $F=f_1\sigma_1\wedge\sigma_2+f_2\sigma_1\wedge\vartheta^{12}+\sigma_2\wedge (f_3\vartheta^{12}+f_4\vartheta^{13})$ but the Einstein equation in the flat directions forces $f_4=0$ leaving no flux on the third flat direction. For $S^2\times T^5$ we have $F=\sigma\wedge(f_1\vartheta^{12}+f_2\vartheta^{34})+\alpha$ with $\alpha\in\Omega^4(T^5)$. Without loss of generality we have either $*\alpha=\vartheta^1$ or $*\alpha=\vartheta^5$ but the latter case is ruled out since the fifth torus direction would not have flux. We must therefore have $F=\sigma\wedge(f_1\vartheta^{12}+f_2\vartheta^{34})+f_3\vartheta^{2345}$ but the Einstein equation in the (12)-directions force $f_3=0$ giving again no flux on the 5-direction.

\subsubsection*{$\fakebold{AdS_3}$ solutions}
With eight-dimensional irreducible factors such as $\mathbbm{CP}^4$ it is not possible to satisfy $F\wedge F=0$. With a six-dimensional factor we can have $\mathbbm{CP}^3\times T^2$ or $\mathbbm{CP}^3\times S^2$ with $F=f\omega\wedge\rho$ or the same with $G_{\mathbbm R}^+(2,5)$ instead of $\mathbbm{CP}^3$. However the Einstein equation forces the six-dimensional factor to be Ricci-flat giving a contradiction. For geometries with an $S^4$ factor $F\wedge F=0$ cannot be satisfied. Again $\mathbbm{CP}^2$ can be treated as a special case of geometries with $S^2$ factors. The geometry $S^3\times S^3\times T^2$ is ruled out since it cannot have flux on all directions while $S^3\times S^2\times S^2\times S^1$ is ruled out since $F\wedge F=0$ cannot be satisfied. Similarly $S^3\times S^2\times T^3$ always has a flat direction without flux and so is also ruled out. For $S^3\times T^5$ we must have $F=f_1\sigma\wedge\vartheta^1+f_2\vartheta^1\wedge\alpha$ with $\alpha\in\Omega^3(T^4)$ but we can take $*\alpha=\vartheta^5$ without loss of generality and then there is no flux on the 5-direction giving again a contradiction.

This leaves geometries with $S^2$ factors and flat directions (recall that $\mathbbm{CP}^2$ is treated as a special of these). First we have the possibilities with three or four $S^2$ factors
\begin{align}
&\fakebold{AdS_3\times\underline{S^2\times S^2\times S^2(H^2)\times S^2(H^2)}}\,,\qquad
\fakebold{AdS_3\times\underline{S^2\times S^2\times S^2(H^2)\times T^2}}\,,
\nonumber\\
&
F=f_{ij}\sigma_i\wedge\sigma_j\,,\qquad\mbox{with}\qquad f_{12}f_{34}+f_{13}f_{24}+f_{14}f_{23}=0\,,
%
%
%
\label{eq:F-26}
\end{align}
where $\sigma_4$ is either the volume form of the last $S^2$ or $\vartheta^{12}$. The curvature of the first two-dimensional factor is proportional to $2f_{12}^2+2f_{13}^2+2f_{14}^2-f_{23}^2-f_{24}^2-f_{34}^2$ and the remaining ones are obtained by $1\leftrightarrow2$ etc. with the last one vanishing in the $T^2$ case. The sum of any three is positive which gives the above possibilities.

For $AdS_3\times S^2\times S^2\times T^4$ the flux takes the form
\begin{equation}
F=f_1\sigma_1\wedge\sigma_2+\sigma_1\wedge(f_2\vartheta^{12}+f_3\vartheta^{34})+\tfrac12f_{ij}\sigma_2\wedge\vartheta^{ij}+f_4\vartheta^{1234}\,,
\end{equation}
where we take, without loss of generality, $f_{14}=0$. The condition $F\wedge F=0$ implies $f_1f_4+f_2f_{34}+f_3f_{12}=0$ and the Einstein equation in the torus directions implies $f_{23}=0$ and
\begin{equation}
f_{24}=\pm f_{13}\,,\qquad
f_2^2=f_3^2-f_{12}^2+f_{34}^2\,,\qquad
f_1^2=f_3^2+2f_4^2+f_{13}^2+f_{34}^2\,,\qquad
f_{13}(f_{12}\pm f_{34})=0\,.
\end{equation}
We find the following three solutions
\begin{align}
(i)\qquad&
f_{13}=f_{24}=0\,,\qquad
f_1^2=\tfrac12\Big(f_3^2+f_{34}^2+\sqrt{(f_3^2+f_{34}^2)^2+8(f_2f_{34}+f_3f_{12})^2}\Big)\,,
\nonumber\\
&f_2^2=f_3^2-f_{12}^2+f_{34}^2\,,\qquad
f_4=-(f_2f_{34}+f_3f_{12})/f_1\,,
\\
(ii)\qquad&
f_4=0\,,\qquad
f_{24}=f_{13}\,,\qquad
f_{34}=-f_{12}\,,\qquad
f_3=f_2\,,\qquad
f_1^2=f_2^2+f_{13}^2+f_{12}^2
\\
(iii)\qquad&
f_{24}=-f_{13}\,,\qquad
f_{34}=f_{12}\,,\qquad
f_3=f_2\,,\qquad
f_4=-2f_2f_{12}/f_1\,,\qquad
\nonumber\\
&f_1^2=\tfrac12\Big(f_2^2+f_{13}^2+f_{12}^2+\sqrt{(f_2^2+f_{13}^2+f_{12}^2)^2+32f_2^2f_{12}^2}\Big)\,.
\end{align}
The first solution gives
\begin{align}
\fakebold{AdS_3\times\underline{S^2\times S^2\times T^4}}\,,\qquad 
F=&2f_1\sigma_1\wedge\sigma_2+\sigma_1\wedge(f_2(\omega_++\omega_-)+f_3(\omega_+-\omega_-))
\nonumber\\
&{}
+\sigma_2\wedge(f_{12}(\omega_++\omega_-)+f_{34}(\omega_+-\omega_-))
+f_4\omega_+\wedge\omega_+\,,
\label{eq:F-28-1}
%
%
\end{align}
with $f_1^2=\tfrac12\Big(f_3^2+f_{34}^2+\sqrt{(f_3^2+f_{34}^2)^2+8(f_2f_{34}+f_3f_{12})^2}\Big)$, $f_2^2=f_3^2-f_{12}^2+f_{34}^2$ and $f_4=-(f_2f_{34}+f_3f_{12})/f_1$. The curvatures of the two $S^2$'s are proportional to $f_2^2+f_3^2+f_4^2$ and $f_4^2+f_{12}^2+f_{34}^2$ respectively. The second solution gives
\begin{align}
\fakebold{AdS_3\times\underline{S^2\times S^2\times T^4}}\,,\qquad F=f_1\sigma_1\wedge\sigma_2+f_2\sigma_1\wedge\omega_+
+\sigma_2\wedge(f_{12}\omega_-+f_{13}\mathrm{Re}\,\Omega_-)
\,,
\label{eq:F-28-2}
%
%
%
\end{align}
with $f_1^2=f_2^2+f_{13}^2+f_{12}^2$. The curvatures of the two $S^2$'s are proportional to $f_2^2$ and $f_{12}^2+f_{13}^2$ respectively. Finally the third solution gives
\begin{align}
\fakebold{AdS_3\times\underline{S^2\times S^2\times T^4}}\,,\qquad F=&f_1\sigma_1\wedge\sigma_2+f_2\sigma_1\wedge\omega_+
+\sigma_2\wedge(f_{12}\omega_++f_{13}\mathrm{Re}\,\Omega_+)
\nonumber\\
&{}
-\frac{f_2f_{12}}{f_1}\omega_+\wedge\omega_+\,,
\label{eq:F-28-3}
%
%
%
\end{align}
with $f_1^2=\frac12\Big(f_2^2+f_{13}^2+f_{12}^2+\sqrt{(f_2^2+f_{13}^2+f_{12}^2)^2+32f_2^2f_{12}^2}\Big)$. The curvatures of the two $S^2$'s are proportional to $2f_2^2(1+2f_{12}^2/f_1^2)$ and $2f_{12}^2(1+2f_2^2/f_1^2)+2f_{13}^2$ respectively. 

Solutions with $\mathbbm{CP}^2$ factors are obtained as special cases of the above analysis. With two $\mathbbm{CP}^2$ factors we find the solutions\footnote{These were missed in \cite{FigueroaO'Farrill:2011fj}.}
\begin{equation}
\fakebold{AdS_3\times\underline{\mathbbm{CP}^2\times\mathbbm{CP}^2(\mathbbm{CH}^2)}}\,,\qquad
F=\tfrac12f_1^2\omega_1^2+\tfrac{1}{\sqrt2}f_1f_4\omega_1\wedge\omega_2-\tfrac12f_4^2\omega_2^2\,,
%
%
\label{eq:F-22}
\end{equation}
where the curvatures of the four-dimensional factors are proportional to $(f_1^2+f_4^2)^2+f_1^4-2f_4^4$ and $(f_1^2+f_4^2)^2+f_4^4-2f_1^4$, respectively, so that at most one can have negative curvature. With one $\mathbbm{CP}^2$ factor we find the solutions
\begin{align}
&\fakebold{AdS_3\times\underline{\mathbbm{CP}^2\times S^2(H^2)\times S^2(H^2)}}\,,\qquad
\fakebold{AdS_3\times\underline{\mathbbm{CH}^2\times S^2\times S^2}}\,,\nonumber\\
&\qquad F=\tfrac12f_1\omega^2+f_2\omega\wedge\sigma_1+f_3\omega\wedge\sigma_2+f_4\sigma_1\wedge\sigma_2\,,\qquad f_1f_4+2f_2f_3=0\,.
%
%
%
\label{eq:F-23}
\end{align}
The curvature of $\mathbbm{CP}^2$ is proportional to $2f_1^2+f_2^2+f_3^3-f_4^2$ while those of the two-dimensional factors are proportional to $-f_1^2+4f_2^2-2f_3^2+2f_4^2$ and $-f_1^2-2f_2^2+4f_3^2+2f_4^2$ respectively. The sum is positive and if the first is negative the latter two are positive definite giving the solutions listed above. With some directions flat we find the solutions
\begin{equation}
\fakebold{AdS_3\times\underline{\mathbbm{CP}^2\times S^2(H^2)\times T^2}}\,,\qquad
\fakebold{AdS_3\times\underline{\mathbbm{CP}^2\times T^4}}\,,\qquad
F=\tfrac12f_1\omega\wedge(\omega-\vartheta^{12})+f_2\sigma\wedge(\omega+\vartheta^{12})\,,
%
%
%
\end{equation}
which are special cases of the above solutions. In the $T^4$ case we take $\sigma=\vartheta^{34}$ and $f_2=-\frac12f_1$.

Finally for $AdS_3\times S^2\times T^6$ the flux takes the form $F=\sigma\wedge\alpha+\beta$ with $\alpha\in\Omega^2(T^6)$ and $\beta\in\Omega^4(T^6)$. The trace of the Einstein equation in the torus directions implies
\begin{equation}
0=|\alpha|^2+2|\beta|^2-|F|^2=|\beta|^2\,,
\end{equation}
so that $\beta$ vanishes. Taking $\alpha=f_1\vartheta^{12}+f_2\vartheta^{34}+f_3\vartheta^{56}$ the Einstein equation implies $f_1=f_2=f_3$ (up to irrelevant signs) and we find the solution
\begin{equation}
\fakebold{AdS_3\times\underline{S^2\times T^6}}\,,\qquad F=f\sigma\wedge\omega\,.
\label{eq:F-29}
\end{equation}
This solution can be obtained by setting $f_2=0$ in (\ref{eq:F-28-3}).

\subsubsection*{$\fakebold{AdS_2}$ solutions}
With a nine-dimensional irreducible factor we get the solution
\begin{equation}
\fakebold{AdS_2\times\underline{SLAG_4}}\,,\qquad F=f\Omega\,.
\end{equation}
The only eight-dimensional irreducible factor with an invariant 3-form is $SU(3)$ so we can have flux on $SU(3)\times S^1$ of the form $F=f\Omega\wedge\vartheta^1$ but the Einstein equation in the flat direction forces $|F|^2=0$ ruling this case out. With a six-dimensional irreducible factor it is not possible to put flux on the remaining three directions. There are no suitable five-dimensional irreducible spaces and $S^4$ is ruled out by $F\wedge F=0$. $\mathbbm{CP}^2$ is ruled out as a special case of the analysis for $S^2$ factors. Solutions with $S^3$ factors are ruled out by the same analysis as in the $|F|^2=0$ case since we did not use the condition $|F|^2=0$ there. Finally, backgrounds with only $S^2$ factors and flat directions can't have flux on all flat directions as we prove in appendix \ref{app:ads2-s2}.

\subsection{Solutions with $|F|^2\neq0$: single Riemannian factor without flux}\label{sec:no-S-flux}
Here we consider solutions with $|F|^2\neq0$ and where precisely one indecomposable Riemannian factor is without flux.

\subsubsection*{$\fakebold{AdS_4}$ solutions}
Since the flux involves a term proportional to the volume form on $AdS_4$ the only way to satisfy $F\wedge F=0$ is to have no flux on the remaining directions which must then form an $S^7$
\begin{equation}
\fakebold{\underline{AdS_4}\times S^7}\qquad F=f\nu\,.
\end{equation}

\subsubsection*{$\fakebold{AdS_3}$ solutions}
As seen in sec. \ref{sec:F2-zero} solutions with flux on $AdS_3$ must have $|F|^2=0$.

\subsubsection*{$\fakebold{AdS_2}$ solutions}
With an seven-dimensional factor without flux we have the solution
\begin{equation}
\fakebold{\underline{AdS_2\times H^2}\times S^7}\,,\qquad F=f\nu\wedge\sigma\,.
\end{equation}
From the Einstein equation the curvatures are $-\frac{f^2}{3}$, $-\frac{f^2}{3}$ and $\frac{f^2}{6}$ respectively. The next possibility is a five-dimensional factor without flux. We find
\begin{align}
&\fakebold{\underline{AdS_2\times S^2(H^2)\times S^2(H^2)}\times SLAG_3}\,,\qquad
\fakebold{\underline{AdS_2\times S^2\times S^2}\times SL(3,\mathbbm R)/SO(3)}\,,
\nonumber\\
&\fakebold{\underline{AdS_2\times S^2(H^2)\times S^2(H^2)}\times S^5}\,,\qquad
\fakebold{\underline{AdS_2\times S^2\times S^2}\times H^5}\,,
\nonumber\\
&\qquad F=f_1\nu\wedge\sigma_1+f_2\nu\wedge\sigma_2+f_3\sigma_1\wedge\sigma_2\,,
\label{eq:F-50,51}
\end{align}
where $SL(3,\mathbbm R)/SO(3)$ is the non-compact version of $SLAG_3$. From the Einstein equation the curvatures are $\frac16$ times, respectively,  $-2f_1^2-2f_2^2-f_3^2$, $-2f_1^2+f_2^2+2f_3^2$, $f_1^2-2f_2^2+2f_3^2$ and $f_1^2+f_2^2-f_3^2$. The sum of last three is non-negative and if the last one is negative the other two are positive giving the possibilities listed. When one of the curvatures vanish we get the solutions
\begin{align}
\fakebold{\underline{AdS_2\times S^2(H^2)\times T^2}\times S^5}\,,\qquad
\fakebold{\underline{AdS_2\times S^2(H^2)\times T^2}\times SLAG_3}\,,
\nonumber\\
F=f_1\nu\wedge\sigma+f_2\nu\wedge\vartheta^{12}+f_3\sigma\wedge\vartheta^{12}\,,\qquad f_2^2=\tfrac12f_1^2+f_3^2\,,
\label{eq:F-52,53}
\end{align}
with curvatures $-\frac14f_1^2-\frac12f_2^2$, $-\frac12f_1^2+\frac12f_2^2$ and $\frac14f_1^2$ respectively. And when the curvature of two factors vanish we find
\begin{equation}
\fakebold{\underline{AdS_2\times T^4}\times S^5}\,,\qquad
\fakebold{\underline{AdS_2\times T^4}\times SLAG_3}\,,\qquad
F=f\nu\wedge\omega_++\tfrac{1}{2\sqrt2}f\omega_+\wedge\omega_+\,,
\label{eq:F-54,55}
\end{equation}
with curvatures $-\frac34f^2$ and $\frac14f^2$. It it easy to see that this gives the most general solution of this form. Another special case of this analysis gives the solutions
\begin{align}
&\fakebold{\underline{AdS_2\times\mathbbm{CP}^2(\mathbbm{CH}^2)}\times SLAG_3}\,,\qquad
\fakebold{\underline{AdS_2\times\mathbbm{CP}^2}\times SL(3,\mathbbm R)/SO(3)}\,,\qquad
\nonumber\\
&
\fakebold{\underline{AdS_2\times\mathbbm{CP}^2(\mathbbm{CH}^2)}\times S^5}\,,\qquad
\fakebold{\underline{AdS_2\times\mathbbm{CP}^2}\times H^5}\,,\qquad
F=f_1\nu\wedge\omega+\tfrac12f_2\omega^2\,.
\label{eq:F-48,49}
\end{align}
The curvatures are ($\frac16$ times) $-4f_1^2-f_2^2$,  $-f_1^2+2f_2^2$ and $2f_1^2-f_2^2$.

With a four-dimensional factor without flux it is not possible to put flux on all the remaining directions (in non-trivial examples this follows as a special case of the analysis in appendix \ref{app:ads2-s2}). Next we consider the case of an $S^3$ factor without flux. With an irreducible six-dimensional manifold such as $\mathbbm{CP}^3$ or $G^+_{\mathbbm R}(2,5)$ the flux takes the form $F=f_1\nu\wedge\omega+f_2\omega^2$ but $F\wedge F=0$ forces $f_2=0$ and then the curvature of the six-manifold vanishes giving a contradiction. With flux on an $S^4$ factor it is not possible to satisfy $F\wedge F=0$. The $\mathbbm{CP}^2$ case is treated below. We can have at most one $S^3$ factor with flux but it is not possible to satisfy $F\wedge F=0$ and have flux on all remaining directions.

This leaves only $S^2$ factors and flat directions with flux. With $S^2\times S^2\times S^2$ we find the solutions
\begin{align}
&\fakebold{\underline{AdS_2\times S^2\times S^2(H^2)\times S^2(H^2)}\times S^3}\,,\qquad
\fakebold{\underline{AdS_2\times S^2\times S^2\times S^2(H^2)}\times H^3}\,,
\label{eq:F-57}
\\
&F=f_1(\nu\wedge\sigma_1+f_2\nu\wedge\sigma_2+f_3\nu\wedge\sigma_3)+f_4\sigma_1\wedge\sigma_2+f_5\sigma_1\wedge\sigma_3-(f_2f_5+f_3f_4)\sigma_2\wedge\sigma_3\,.
\nonumber
\end{align}
The curvatures are proportional to $-2f_1^2(1+f_2^2+f_3^2)-f_4^2-f_5^2-(f_2f_5+f_3f_4)^2$, $f_1^2(-2+f_2^2+f_3^2)+2f_4^2+2f_5^2-(f_2f_5+f_3f_4)^2$, $f_1^2(1-2f_2^2+f_3^2)+2f_4^2-f_5^2+2(f_2f_5+f_3f_4)^2$, $f_1^2(1+f_2^2-2f_3^2)-f_4^2+2f_5^2+2(f_2f_5+f_3f_4)^2$ and $f_1^2(1+f_2^2+f_3^2)-f_4^2-f_5^2-(f_2f_5+f_3f_4)^2$, respectively. The sum of any three of the last four is non-negative giving the possibilities listed. The special case with one curvature vanishing gives
\begin{align}
&\fakebold{\underline{AdS_2\times S^2\times S^2(H^2)\times T^2}\times S^3}\,,\qquad
\fakebold{\underline{AdS_2\times S^2\times S^2\times T^2}\times H^3}\,,\nonumber\\
&\qquad F=f_1(\nu\wedge\sigma_1+f_2\nu\wedge\sigma_2+f_3\nu\wedge\vartheta^{12})+f_4\sigma_1\wedge\sigma_2+f_5\sigma_1\wedge\vartheta^{12}-(f_2f_5+f_3f_4)\sigma_2\wedge\vartheta^{12}\,,
\nonumber\\
&
f_2=\Big(-2f_3f_4f_5\pm\sqrt{(f_1^2(1-2f_3^2)+2f_5^2)(f_4^2-f_1^2-2f_5^2)}\Big)/(f_1^2+2f_5^2)\,.
%
%
\end{align}
The curvatures are proportional to $-f_1^2(1+f_2^2)-f_5^2-(f_2f_5+f_3f_4)^2$, $f_1^2(f_2^2-f_3^2)+2f_5^2+(f_2f_5+f_3f_4)^2$, $f_1^2(1-f_3^2)+f_5^2+2(f_2f_5+f_3f_4)^2$ and $f_1^2f_3^2-f_5^2-(f_2f_5+f_3f_4)^2$. In the limit where the curvature of the $S^3$ factor goes to zero we obtain the solutions in (\ref{eq:F-41}) and (\ref{eq:F-44}).

A special case of the above analysis gives the solutions with $\mathbbm{CP}^2$ factors
\begin{align}
&\fakebold{\underline{AdS_2\times\mathbbm{CP}^2\times S^2(H^2)}\times S^3(H^3)}\,,\qquad
\fakebold{\underline{AdS_2\times\mathbbm{CH}^2\times S^2}\times S^3}\,,\nonumber\\
&\qquad F=f_1\nu\wedge\sigma+f_2\nu\wedge\omega+f_3\sigma\wedge\omega+f_4\omega^2\,,\qquad
f_1f_4+f_2f_3=0\,.
\label{eq:F-56}
\end{align}
The curvatures are proportional to $-2(f_1^2+2f_2^2+f_3^2+2f_4^2)$, $f_1^2-f_2^2+f_3^2+8f_4^2$, $-2(f_1^2-f_2^2-2f_3^2+2f_4^2)$ and $f_1^2+2f_2^2-2f_3^2-4f_4^2$, respectively. Analyzing the possible signs of the curvatures we find the possibilities listed above. When the curvature of the $S^3$ vanishes we obtain the solution (\ref{eq:F-42}). When the curvature of the $S^2(H^2)$ vanishes we get the solution
\begin{equation}
\fakebold{\underline{AdS_2\times\mathbbm{CP}^2\times T^2}\times S^3(H^3)}\,,\qquad
F=f_1\nu\wedge\vartheta^{12}+f_1\nu\wedge\omega+f_3\vartheta^{12}\wedge\omega-f_3\omega^2\,.
\end{equation}
The curvatures are proportional to $-6(f_1^2+f_3^2)$, $9f_3^2$, and $3(f_1^2-2f_3^2)$. 

In the case where the flux is on $AdS_2\times S^2\times T^4$ the flux takes the form $F=f_1\nu\wedge\sigma+\nu\wedge(f_2\vartheta^{12}+f_3\vartheta^{34})+\frac12f_{ij}\sigma\wedge\vartheta^{ij}+f_4\vartheta^{1234}$ where we can take $f_{14}=0$ without loss of generality. The condition $F\wedge F=0$ leads to $f_1f_4+f_2f_{34}+f_3f_{12}=0$ while the Einstein equation in the torus directions leads to $f_{23}=0$, $f_{24}=\pm f_{13}$ and
\begin{equation}
f_{13}(f_{12}\pm f_{34})=0\,,\qquad
f_2^2=f_1^2+2f_4^2+f_{12}^2+f_{13}^2\,,\qquad
f_3^2=f_1^2+2f_4^2+f_{13}^2+f_{34}^2\,.
\end{equation}
It is easy to see that $f_2\neq0$ since otherwise we cannot have flux on all torus directions.  We find the following five solutions
\begin{align}
(i)\qquad&
f_{24}=f_{13}=0\,,\qquad
f_{34}=-f_1f_4/f_2-f_3f_{12}/f_2\,,\qquad
f_2^2=f_1^2+2f_4^2+f_{12}^2\,,\qquad
\nonumber\\
&f_3=\Big(f_1f_4f_{12}\pm f_2\sqrt{(f_1^2+2f_4^2)^2+f_1^2f_4^2}\Big)/(f_1^2+2f_4^2)
\nonumber\\
(ii)\qquad&
f_{24}=f_{13}=0\,,\qquad
f_{12}=f_2\,,\qquad
f_{34}=-f_3\,,\qquad
f_1=f_4=0
\nonumber\\
(iii)\qquad&
f_{24}=f_{13}\,,\qquad
f_{34}=-f_{12}\,,\qquad
f_3=f_2\,,\qquad
f_1=0\,,\qquad
f_2^2=2f_4^2+f_{12}^2+f_{13}^2
\nonumber\\
(iv)\qquad&
f_{24}=f_{13}\,,\qquad
f_{34}=-f_{12}\,,\qquad
f_3=f_2\,,\qquad
f_4=0\,,\qquad
f_2^2=f_1^2+f_{12}^2+f_{13}^2
\nonumber\\
(v)\qquad&
f_{24}=-f_{13}\,,\qquad
f_{34}=f_{12}=-f_1f_4/2f_2\,,\qquad
f_3=f_2\,,\qquad
\nonumber\\
&f_2^2=\tfrac12(f_1^2+2f_4^2+f_{13}^2)+\tfrac12\sqrt{(f_1^2+2f_4^2+f_{13}^2)^2+f_1^2f_4^2}\,.
\end{align}
Case (ii) is ruled out since it has $|F|^2=0$ and we have the following solutions.
\begin{align}
\fakebold{\underline{AdS_2\times S^2\times T^4}\times S^3}\,,\qquad
F=&
2f_1\nu\wedge\sigma
+\nu\wedge(f_2(\omega_++\omega_-)+f_3(\omega_+-\omega_-))
\nonumber\\
&{}
+\sigma\wedge(f_{12}(\omega_++\omega_-)+f_{34}(\omega_+-\omega_-))
+f_4\omega_+\wedge\omega_+\,,
\label{eq:F-60-1}
\end{align}
with $f_{34}=-(f_1f_4+f_3f_{12})/f_2$, $f_3=\Big(f_1f_4f_{12}\pm f_2\sqrt{(f_1^2+2f_4^2)^2+f_1^2f_4^2}\Big)/(f_1^2+2f_4^2)$ and $f_2^2=f_1^2+2f_4^2+f_{12}^2$. The Ricci tensors are proportional to $-(2f_1^2+3f_4^2+f_{12}^2+f_{34}^2)$, $f_4^2+f_{12}^2+f_{34}^2$ and $f_1^2+f_4^2$ respectively. In the limit $f_1,f_4\rightarrow 0$ the curvature of the $S^3$ factor goes to zero and we obtain the $AdS_2\times S^2\times T^7$ solution in (\ref{eq:F-45-1}).
\begin{equation}
\fakebold{\underline{AdS_2\times S^2\times T^4}\times S^3}\,,\qquad
F=f_2\nu\wedge\omega_++\sigma\wedge(f_{12}\omega_-+f_{13}\mathrm{Re}\,\Omega_-)+\tfrac12f_4\omega_+\wedge\omega_+\,,
\label{eq:F-60-2}
\end{equation}
with $f_2^2=2f_4^2+f_{12}^2+f_{13}^2$. The Ricci tensors are proportional to $-(3f_4^2+2f_{12}^2+2f_{13}^2)$, $f_4^2+2f_{12}^2+2f_{13}^2$ and $f_4^2$ respectively. In the limit $f_4\rightarrow 0$ the curvature of the $S^3$ factor goes to zero and we obtain the $AdS_2\times S^2\times T^7$ solution in (\ref{eq:F-45-2}).
\begin{equation}
\fakebold{\underline{AdS_2\times S^2\times T^4}\times S^3}\,,\qquad
F=f_1\nu\wedge\sigma+f_2\nu\wedge\omega_++\sigma\wedge(f_{12}\omega_-+f_{13}\mathrm{Re}\,\Omega_-)\,,
\label{eq:F-60-3}
\end{equation}
with $f_2^2=f_1^2+f_{12}^2+f_{13}^2$. The Ricci tensors are proportional to $-f_1^2-f_{12}^2-f_{13}^2$, $f_{12}^2+f_{13}^2$ and $f_1^2$ respectively. In the limit $f_1\rightarrow 0$ the curvature of the $S^3$ factor goes to zero and we obtain the $AdS_2\times S^2\times T^7$ solution in (\ref{eq:F-45-2}).
\begin{equation}
\fakebold{\underline{AdS_2\times S^2\times T^4}\times S^3}\,,\qquad
F=f_1\nu\wedge\sigma+f_2\nu\wedge\omega_++\sigma\wedge(f_{12}\omega_++f_{13}\mathrm{Re}\,\Omega_+)+\tfrac12f_4\omega_+\wedge\omega_+\,,
\label{eq:F-60-4}
\end{equation}
with $f_{12}=-f_1f_4/2f_2$ and $f_2^2=\tfrac12\Big(f_1^2+2f_4^2+f_{13}^2+\sqrt{(f_1^2+2f_4^2+f_{13}^2)^2+f_1^2f_4^2}\Big)$. The Ricci tensors are proportional to $-(2f_1^2+3f_4^2+2f_{12}^2+2f_{13}^2)$, $f_4^2+2f_{12}^2+2f_{13}^2$ and $f_1^2+f_4^2$ respectively. In the limit $f_1,f_4\rightarrow 0$ the curvature of the $S^3$ factor goes to zero and we obtain the $AdS_2\times S^2\times T^7$ solution in (\ref{eq:F-45-3}).

With flux on $AdS_2\times T^6$ we have $F=\nu\wedge\alpha+\beta$ for $\alpha\in\Omega^2(T^6)$ and $\beta\in\Omega^4(T^6)$. Taking the trace of the Einstein equation in the torus directions we find $|\beta|^2=0$ so that $\beta=0$. Taking $\alpha=f_1\vartheta^{12}+f_2\vartheta^{34}+f_3\vartheta^{56}$ the Einstein equation in the torus directions forces $f_3=f_2=f_1$ (up to irrelevant signs) and we find the solution
\begin{equation}
\fakebold{\underline{AdS_2\times T^6}\times S^3}\,,\qquad
F=f_1\nu\wedge\omega\,,
\label{eq:F-61}
\end{equation}
where the curvatures are proportional to $-f_1^2$ and $\frac12f_1^2$ respectively.

It remains to analyze backgrounds with an $S^2$ factor without flux. With an irreducible six-dimensional factor it is not possible to put flux on all directions. We can therefore only have $S^3$ or $S^2$ factors (or $\mathbbm{CP}^2$ which we treat as a special case of $S^2\times S^2$). We can at most have one $S^3$ factor. Taking $S^3\times S^2\times T^2$ it is easy to see that the Einstein equation in the torus directions cannot be satisfied. The same is true for $S^3\times T^4$. This leaves only backgrounds with $S^2$ factors (or $\mathbbm{CP}^2$ factors) and flat directions but these are ruled out by the analysis in appendix \ref{app:ads2-s2}.

\subsection{Solutions with $|F|^2\neq0$: remaining cases}\label{sec:remaining}
Let us first note that it is not possible to have flux on all directions. It is clear that this is impossible for $AdS_n$ solutions with $n>3$. For $n=3$ it follows from the analysis in sec. \ref{sec:F2-zero} while for $n=2$ it follows from appendix \ref{app:ads2-s2}. Therefore the remaining solutions must have at least two indecomposable Riemannian factors without flux.

We will not include here solutions which are special cases of ones already constructed where parameters can be chosen so that the flux is on fewer directions but the geometry remains unchanged, e.g. $AdS_5\times\underline{S^2\times S^2}\times H^2$ obtained by setting $f_2=f_3=0$ in (\ref{eq:F-8}).

All $AdS_7$ solutions are already accounted for. There can not be any additional $AdS_6$ solutions either since a flat direction without flux implies $|F|^2=0$.

\subsubsection*{$\fakebold{AdS_5}$ solutions}
Since there is no flux on the AdS part we have $|F|^2>0$ so that the remaining directions without flux must form a space of negative curvature. Ignoring special cases of previous solutions the only possibility is
\begin{equation}
\fakebold{AdS_5\times\underline{S^4}\times H^2}\,.
\end{equation}
Spaces involving flat directions, e.g. $S^3\times S^1$ or $S^2\times T^2$, are ruled out by the Einstein equation.

\subsubsection*{$\fakebold{AdS_4}$ solutions}
If there is no flux on AdS we again have the solutions with flux on a four-dimensional Riemannian space
\begin{equation}
\fakebold{AdS_4\times\underline{\mathcal M_4}\times H^3}\,,\qquad\mathcal M_4=S^4,\,\mathbbm{CP}^2,\,S^2\times S^2\,.
\end{equation}
If there is flux on AdS there cannot be flux on the other directions because of $F\wedge F=0$. This gives the solutions
\begin{align}
\fakebold{\underline{AdS_4}\times S^5\times S^2}\,,\qquad
\fakebold{\underline{AdS_4}\times SLAG_3\times S^2}\,,\qquad
\fakebold{\underline{AdS_4}\times\mathcal M_4\times S^3}\,.
\end{align}

\subsubsection*{$\fakebold{AdS_3}$ solutions}
As seen in sec. \ref{sec:F2-zero} solutions with flux on $AdS_3$ must have $|F|^2=0$. Therefore there cannot be flux on $AdS_3$ and we again have the solutions
\begin{equation}
\fakebold{AdS_3\times\underline{\mathcal M_4}\times H^4}\,,\qquad\mathcal M_4=S^4,\,\mathbbm{CP}^2,\,S^2\times S^2\,,
\end{equation}
but now we can also replace $H^4$ by $\mathbbm{CH}^2$ or $H^2\times H^2$ giving the solutions (recall that we are not including special cases of previous solutions)
\begin{align}
\fakebold{AdS_3\times\underline{S^4}\times\mathbbm{CH}^2}\,,\qquad
\fakebold{AdS_3\times\underline{S^4}\times H^2\times H^2}\,.
\end{align}
It is not possible to have flux on a five-dimensional space. With flux on six dimensions we have the solutions
\begin{equation}
\fakebold{AdS_3\times\underline{\mathbbm{CP}^3}\times H^2}\,,\qquad
\fakebold{AdS_3\times\underline{G_{\mathbbm R}^+(2,5)}\times H^2}\,,\qquad
\qquad F=f\omega\wedge\omega\,.
\end{equation}
All remaining solutions are special cases of previous ones.

\subsubsection*{$\fakebold{AdS_2}$ solutions}
Let us first analyze the case without flux on AdS. We have again the solutions with flux on $S^4$ (other four-dimensional spaces are special cases of more general backgrounds)
\begin{equation}
\fakebold{AdS_2\times\underline{S^4}\times H^5}\,,\qquad
\fakebold{AdS_2\times\underline{S^4}\times SL(3,\mathbbm R)/SO(3)}\,,\qquad
\fakebold{AdS_2\times\underline{S^4}\times H^3\times H^2}\,.
\end{equation}
The next possibility is to have flux on six directions and, just as in the $AdS_3$ case, we find the solutions
\begin{equation}
\fakebold{AdS_2\times\underline{\mathbbm{CP}^3}\times H^3}\,,\qquad
\fakebold{AdS_2\times\underline{G_{\mathbbm R}^+(2,5)}\times H^3}\,,\qquad
\qquad F=f\omega\wedge\omega\,.
\end{equation}
The rest are special cases of previous solutions. The case with flux on seven directions is ruled out since the flux must take the form $F=\rho\wedge\vartheta^1$ for some three-form $\rho$ but the Einstein equation in the 1-direction forces $|F|^2=0$. 

For backgrounds with flux on AdS we have the solution
\begin{equation}
\fakebold{\underline{AdS_2\times H^2}\times S^4\times S^3}\,,\qquad
F=f\nu\wedge\sigma\,.
\end{equation}
All remaining possibilities are either special cases of previous backgrounds or ruled out by the analysis in appendix \ref{app:ads2-s2}.

\section{Supersymmetry analysis}\label{sec:susy}
Having found all symmetric space solutions of eleven-dimensional supergravity we now turn to the question of which ones preserve some supersymmetry. To do this we will analyze the integrability condition for the Killing spinor equation which arises by requiring the supersymmetry variation of the gravitino field strength to vanish or $\xi^\alpha\nabla_\alpha\psi_{ab}^\beta|_{\theta=0}=0$, where $\xi$ is the Killing spinor. Using the supergravity constraint (\ref{eq:Uab}) and restricting to the symmetric space case so that all fields are constant this condition becomes
\begin{equation}
U_{ab}\xi=(2G_{[a}G_{b]}-\tfrac14R_{ab}{}^{cd}\Gamma_{cd})\xi=0\,,\qquad G_a=\tfrac18\slashed F\Gamma_a-\tfrac{1}{24}\Gamma_a\slashed F\,.
\end{equation}
It is convenient to split the tangent space index as $a=(\tilde a,\,\u a)$ where $\tilde a$ runs over directions without flux and $\u a$ runs over directions with flux. The supersymmetry conditions $U_{\tilde a\tilde b}\xi=0$, $U_{\tilde a\u b}\xi=0$ and $U_{\u a \u b}\xi=0$ then simplify by noting that $[\Gamma_{\tilde a},\slashed F]=0$ and they become\footnote{The first two equations are of course not there if there is flux on all directions.}
\begin{align}
&\big(\Gamma_{\tilde a\tilde b}\slashed F^2-18R_{\tilde a\tilde b}{}^{\tilde c\tilde d}\Gamma_{\tilde c\tilde d}\big)\xi=0\,,\qquad
\big(3\slashed F^2\Gamma_{\u b}+2\slashed F\Gamma_{\u b}\slashed F-\Gamma_{\u b}\slashed F^2\big)\xi=0\,,
\label{eq:susy12}
\\
&\big(
9\slashed F\Gamma_{[\u a}\slashed F\Gamma_{\u b]}
-3\slashed F\Gamma_{\u a\u b}\slashed F
-3\Gamma_{[\u a}\slashed F^2\Gamma_{\u b]}
+\Gamma_{[\u a}\slashed F\Gamma_{\u b]}\slashed F
-72R_{\u a\u b}{}^{\u c\u d}\Gamma_{\u c\u d}
\big)\xi=0\,.
\label{eq:susy3}
\end{align}
By contracting with gamma matrices and using the Einstein equation we can derive an extra condition which is sometimes useful\footnote{Using the fact that $\slashed F^2-|F|^2$ only has a $\Gamma_{(4)}$ piece and $\slashed F\Gamma_{\u b}\slashed F-\Gamma_{\u b}|F|^2+2\langle i_{\u b}F,i_{\u c}F\rangle\Gamma^{\u c}$ only has a $\Gamma_{(5)}$ piece one can show that this is the only relation implied.}
\begin{equation}
\big((10-n)\slashed F^2-6|F|^2\big)\xi=0\,.
\label{eq:extra-susy}
\end{equation}
We will now analyze these equations for the symmetric space AdS solutions in tables \ref{tab:ads7}--\ref{tab:ads2}. The results are summarized in table \ref{tab:susy-solutions}. For the corresponding analysis for Cahen-Wallach (pp-wave) solutions we refer to \cite{Cvetic:2002si,Gauntlett:2002cs}.

The first condition in (\ref{eq:susy12}) implies that the part of the geometry without flux cannot be a product of (non-flat) irreducible factors. This follows by noting that if we take $\tilde a$ and $\tilde b$ to belong to different factors the curvature term drops out and we find $\slashed F^2\xi=0$ but by the condition (\ref{eq:extra-susy}) this can only happen if the part without flux is flat since otherwise $|F|^2$ is non-zero. This simple observation rules out supersymmetry for all the solutions found in sec. \ref{sec:remaining}. The ones that remain to analyze are those with either a single irreducible factor without flux or only flat directions without flux. These are listed in table \ref{tab:all-cases} according to the number of directions with flux $n$.\footnote{For some of these backgrounds one can adjust parameters so that the flux is on fewer directions. These cases should in principle be analyzed separately. However, it is clear that the only thing that can happen is that either one gets a solution with more than one non-flat irreducible factor without flux, in which case supersymmetry is already ruled out, or one gets a solution with additional flat directions without flux, in which case the supersymmetry conditions (\ref{eq:susy12}) and (\ref{eq:susy3}) remain unchanged. Therefore these special cases are already covered by our analysis.} Except for some special cases involving flux on $T^4$ or $S^3$ indicated in boldface the flux takes a very simple form: $\slashed F$ consists of a number of terms that all commute with each other and can therefore be simultaneously diagonalized. This makes the supersymmetry analysis quite simple and we will consider these backgrounds first before turning to the five special non-diagonalizable cases indicated in boldface.

\begin{table}[ht]
\begin{tabular}{l|l}
$n$ & Background number (from tables \ref{tab:ads7}--\ref{tab:ads2})\\
\hline 4 & 1, 2, 3, 11, 47\\
\hline 6 & 4, 5, 6, (7), 8, (9), 48, 49, 50, 51, (52, 53, 54, 55)\\
\hline 7 & ({\bf 21})\\
\hline 8 & 22, 23, (24, 25), 26, (27), {\bf 28}, (29), 41, 42, (44, {\bf 45}), 56, 57, (58, 59), {\bf 60}, (61)\\
\hline 9 & {\bf 46}\\
\hline 10 &{\bf 20}, 37, 38, 39, 40, {\bf 43}
\end{tabular}
\caption{AdS backgrounds for which supersymmetry is not trivially ruled out organized according to the number of directions with flux $n$. Those that can be treated as special cases of others in the list are in parenthesis.}
\label{tab:all-cases}
\end{table}

\subsection{Diagonalizable $\slashed F$}
For all the backgrounds we need to analyze in table \ref{tab:all-cases} except the ones in boldface the flux takes the general form (for a suitable numbering of the coordinates)
\begin{equation}
\slashed F=
a_1\Gamma^{0123}
+a_2\Gamma^{0145}
+a_3\Gamma^{0167}
+a_4\Gamma^{0189}
+a_5\Gamma^{2345}
+a_6\Gamma^{2367}
+a_7\Gamma^{2389}
+a_8\Gamma^{4567}
+a_9\Gamma^{4589}
+a_{10}\Gamma^{6789}\,,
\label{eq:F10}
\end{equation}
where $a_1\,,\ldots\,,a_{10}$ are determined by the parameters entering $F$ for the background in question. As already mentioned the important point is that the matrices $\Gamma^{0123}$, $\Gamma^{0145}$, etc. all commute and can therefore be simultaneously diagonalized. Their eigenvalues are respectively $i\epsilon_1,\,\ldots,\,i\epsilon_4$, $\epsilon_5,\,\ldots,\,\epsilon_{10}$ where the $\epsilon$'s are $\pm1$. Since the matrices are not independent we have the following relations between the eigenvalues
\begin{equation}
\epsilon_5=-\epsilon_1\epsilon_2\,,\quad
\epsilon_6=-\epsilon_1\epsilon_3\,,\quad
\epsilon_7=-\epsilon_1\epsilon_4\,,\quad
\epsilon_8=-\epsilon_2\epsilon_3\,,\quad
\epsilon_9=-\epsilon_2\epsilon_4\,,\quad
\epsilon_{10}=-\epsilon_3\epsilon_4\,.
\end{equation}

The first supersymmetry condition in (\ref{eq:susy12}) is not very constraining since it gives only one equation for all $\u a$. The second equation however has $n$ components, one for each value of $\u b$, and turns out to be enough to either rule out supersymmetry or find the supersymmetric solution uniquely in most cases. For a given choice of the free index $\u b$ of the gamma matrix in this equation $\slashed F$ splits up into a piece which commutes with $\Gamma_{\u b}$ and a piece which anti-commutes with it, $\slashed F=\slashed F_1+\slashed F_2$. In terms of these the corresponding equation takes the form
\begin{equation}
(2\slashed F_1^2-\slashed F_1\slashed F_2-3\slashed F_2\slashed F_1)\xi=0\,.
\label{eq:susy2-general}
\end{equation}
For the cases considered here $\slashed F_1$ and $\slashed F_2$ commute so that the equation factorizes into
\begin{equation}
\slashed F_1(\slashed F_1-2\slashed F_2)\xi=0\,.
\end{equation}
Let us start with the cases with flux on ten directions, $n=10$ in table \ref{tab:all-cases}. The second supersymmetry condition in (\ref{eq:susy12}) then gives five equations (the ten choices of $\u b$ are paired)
\begin{align}
(b_5+b_6+b_7+b_8+b_9+b_{10})\big(b_5+b_6+b_7+b_8+b_9+b_{10}-2i(b_1+b_2+b_3+b_4)\big)=&\,0\nonumber\\
(i(b_2+b_3+b_4)+b_8+b_9+b_{10})\big(i(b_2+b_3+b_4)+b_8+b_9+b_{10}-2(ib_1+b_5+b_6+b_7)\big)=&\,0\nonumber\\
(i(b_1+b_3+b_4)+b_6+b_7+b_{10})\big(i(b_1+b_3+b_4)+b_6+b_7+b_{10}-2(ib_2+b_5+b_8+b_9)\big)=&\,0\nonumber\\
(i(b_1+b_2+b_4)+b_5+b_7+b_9)\big(i(b_1+b_2+b_4)+b_5+b_7+b_9-2(ib_3+b_6+b_8+b_{10})\big)=&\,0\nonumber\\
(i(b_1+b_2+b_3)+b_5+b_6+b_8)\big(i(b_1+b_2+b_3)+b_5+b_6+b_8-2(ib_4+b_7+b_9+b_{10})\big)=&\,0\,,
\label{eq:b-eqs}
\end{align}
where we have absorbed the signs from the $\epsilon$'s by defining $b_i=\epsilon_ia_i$. The first equation implies $b_5+b_6+b_7+b_8+b_9+b_{10}=0$. By looking at the most general form of the flux that appears for the $n=10$ backgrounds in table \ref{tab:all-cases} we see that we can take $b_1=b_2=b_3=b_5=b_6=b_8=0$. The remaining equations then imply that $b_4=0$ but from (\ref{eq:F10}) we see that then there is no flux on AdS in contradiction to our assumptions. This rules out all $n=10$ backgrounds in \ref{tab:all-cases} except the ones in boldface.

Next we consider the backgrounds with flux on eight directions, $n=8$ in table \ref{tab:all-cases}. The same analysis applies except that looking at (\ref{eq:F10}) we should either set $a_1=a_2=a_3=a_4=0$ and drop the first equation in (\ref{eq:b-eqs}), corresponding to no flux on AdS, or set $a_4=a_7=a_9=a_{10}=0$ and drop the last equation in (\ref{eq:b-eqs}). In the first case there are four solutions
\begin{align}
(i)&\qquad b_{10}=\tfrac13(2b_5-b_6-b_7)\,,\quad
b_9=\tfrac13(-b_5+2b_6-b_7)\,,\quad
b_8=\tfrac13(-b_5-b_6+2b_7)\,,
\\
(ii)&\qquad b_{10}=2b_5+b_6+b_7\,,\quad
b_9=-b_5-b_7\,,\quad
b_8=b_5+b_6+2b_7\,,
\\
(iii)&\qquad b_{10}=-b_6-b_7\,\quad
b_9=b_5+2b_6+b_7\,,\quad
b_8=b_5+b_6+2b_7\,,
\\
(iv)&\qquad b_{10}=2b_5+b_6+b_7\,,\quad
b_9=b_5+2b_6+b_7\,,\quad
b_8=-b_5-b_6\,.
\end{align}
These can be further reduced by considering the extra supersymmetry condition (\ref{eq:extra-susy}) which reads
\begin{equation}
(b_5+b_6+b_7+b_8+b_9+b_{10})^2-3(b_5^2+b_6^2+b_7^2+b_8^2+b_9^2+b_{10}^2)=0\,.
\end{equation}
Imposing this condition reduces the solutions to
\begin{align}
(i)&\quad b_{10}=b_9=b_8=0\,,\quad
b_7=b_6=b_5\quad
(ii)&
b_{10}=b_8=b_6\,,\quad
b_9=b_7=b_5=0\,\,
\nonumber\\
(iii)&\quad
b_{10}=b_7=b_6=0\,,\quad
b_9=b_8=b_5\quad
(iv)&
b_{10}=b_9=b_7\,,\quad
b_8=b_6=b_5=0\,.
\label{eq:n=8-noAdS}
\end{align}
In the second case, i.e. for backgrounds with flux on AdS, the only solution is
\begin{equation}
b_1=b_2=b_3\,,\qquad
b_4=b_5=b_6=b_7=b_8=b_9=b_{10}=0\,.
\label{eq:n=8-AdS}
\end{equation}
Comparing to the form of $\slashed F$ for the backgrounds with $n=8$ in table \ref{tab:all-cases} we find that background $22$, (\ref{eq:F-22}), is ruled out while $23$ and $26$, (\ref{eq:F-23}) and (\ref{eq:F-26}), reduce to background $29$, (\ref{eq:F-29}), which will be treated below as a special case of background 28. Similarly backgrounds $41$, (\ref{eq:F-41}), and $42$, (\ref{eq:F-42}), are ruled out while $56$ and $57$, (\ref{eq:F-56}) and (\ref{eq:F-57}), are reduced to background 61 which is treated below as a special case of background $60$. 

Next we have the backgrounds with flux on six directions, $n=6$, in table \ref{tab:all-cases}. Looking at (\ref{eq:F10}) we should either set $a_1=a_2=a_3=a_4=a_7=a_9=a_{10}=0$ and drop the first and last equation in (\ref{eq:b-eqs}), corresponding to no flux on AdS, or set $a_3=a_4=a_6=a_7=a_8=a_9=a_{10}=0$ and drop the last two equations in (\ref{eq:b-eqs}). It is easy to see that in both cases there is no non-trivial solution ruling out supersymmetry for all such backgrounds.

For backgrounds with flux on only four directions, i.e. $n=4$ in table \ref{tab:all-cases}, we have $\slashed F^2=|F|^2$ and $\{\Gamma_{\u a},\slashed F\}=0$ and the first two supersymmetry conditions (\ref{eq:susy12}) are trivially satisfied. The last condition (\ref{eq:susy3}) becomes
\begin{equation}
\big(2|F|^2\Gamma_{\u a\u b}+9R_{\u a\u b}{}^{\u c\u d}\Gamma_{\u c\u d}\big)\xi=0\,.
\end{equation}
This rules out backgrounds $2,3$ and $47$ since the curvature does not have the right form.\footnote{For background 2 we have used the fact that the $\mathbbm{CP}^2$ curvature takes the form 
$$
R_{\u a\u b}{}^{\u c\u d}=\frac19|F|^2[-2\delta_{[\u a}^{\u c}\delta_{\u b]}^{\u d}-2\omega_{[\u a}{}^{\u c}\omega_{\u b]}{}^{\u d}+\omega_{\u a\u b}\omega^{\u c\u d}]\,,
$$
where $\omega_{ab}$ is the K\"ahler form.} For backgrounds $1$ and $11$ the curvature of $S^4(AdS_4)$ is $R_{\u a\u b}{}^{\u c\u d}=-\frac29|F|^2\delta_{[\u a}^{\u c}\delta_{\u b]}^{\u d}$ and the above equation is satisfied for any $\xi$, i.e. the backgrounds are maximally supersymmetric. These are the well known Freund-Rubin backgrounds preserving 32 supersymmetries.

\subsection{Non-diagonalizable $\slashed F$}
It remains only to analyze the five backgrounds in boldface in table \ref{tab:all-cases} which are more complicated since $\slashed F$ cannot be completely diagonalized.

\subsubsection*{46. $\fakebold{AdS_2\times SLAG_4}$}
We can easily rule out supersymmetry for this solution. Using the explicit form of the invariant four-form on $SLAG_4$ given in appendix A.3 of \cite{FigueroaO'Farrill:2011fj} we find
\begin{align}
\slashed F=&
\frac{|F|}{\sqrt{18}}
\Big(
\sqrt2(\Gamma^{1249}-\Gamma^{1456}-\Gamma^{1478}-\Gamma^{2579}-\Gamma^{2689})
-\Gamma^{1358}
+\Gamma^{1367}
+\Gamma^{2358}
+\Gamma^{2367}
\nonumber\\
&{}
-\Gamma^{3456}
+\Gamma^{3478}
+\Gamma^{3579}
-\Gamma^{3689}
\Big)\,.
\end{align}
Squaring this expression, noting that the $\sqrt2$-term anti-commutes with the rest, one finds
\begin{equation}
\slashed F^2=|F|^2\,,
\end{equation}
but this is incompatible with the condition in (\ref{eq:extra-susy}) with $n=9$ ruling out supersymmetry for this background.

\subsubsection*{20. $\fakebold{AdS_3\times S^3\times S^3\times T^2}$}
The flux is given by (\ref{eq:F-20}) and we find
\begin{equation}
\slashed F=2f_1\Gamma^{0129}(1-\mathcal P)\,,\qquad\mathcal P=\frac12\left(1-\frac{f_2}{f_1}\Gamma^{012345}-\frac{f_3}{f_1}\Gamma^{012678}\right)\,,\qquad f_1^2=f_2^2+f_3^2\,.
\label{eq:proj-20}
\end{equation}
Note that $\mathcal P$ is a (constant) projection operator projecting on a 16-dimensional subspace of the 32-dimensional space of spinors. Using the fact that the Riemann curvature takes the form $R_{\u a\u b}{}^{\u c\u d}=\frac{c}{2}\delta_{[\u a}^{\u c}\delta_{\u b]}^{\u d}$ where $c=f_1^2,-f_2^2,-f_3^2$ for $AdS_3$ and the first and second $S^3$ respectively it is easy to verify that all the supersymmetry conditions (\ref{eq:susy12}) and (\ref{eq:susy3}) are satisfied provided that the Killing spinor satisfies the projection condition $\mathcal P\xi=\xi$. This background, and the special case where the curvature of one $S^3$ vanishes ($f_3=0$, i.e. solution 21), therefore preserve 16 supersymmetries.


\subsubsection*{28. $\fakebold{AdS_3\times S^2\times S^2\times T^4}$}
The flux takes the form (\ref{eq:F-28-1})--(\ref{eq:F-28-3}) so that
\begin{equation}
\slashed F=
a_1\Gamma^{1234}
+a_2\Gamma^{1256}
+a_3\Gamma^{1278}
+a_4\Gamma^{3456}
+a_5\Gamma^{3478}
+a_6\Gamma^{5678}
+a_7\Gamma^{3457}(1+\pm\Gamma^{5678})\,,
\end{equation}
where $a_1=f_1\neq0$. If $a_7=f_{13}=0$, which includes the first branch (\ref{eq:F-28-1}), $\slashed F$ is diagonalizable and comparing to (\ref{eq:F10}) the solution in (\ref{eq:n=8-noAdS}) gives
\begin{equation}
b_1=b_2=b_3\,,
\end{equation}
(or the equivalent solution $b_1=b_4=b_5$) and the rest vanishing, which reduces the background to 29. This solution turns out to be a special case of the more general solution with $a_7\neq0$ which will be found below. 

It therefore remains to analyze the two branches of solutions with $f_{13}\neq0$, (\ref{eq:F-28-2}) and (\ref{eq:F-28-3}), i.e.
\begin{equation}
\slashed F=
a_1\Gamma^{1234}
+a_2\Gamma^{1256}(1-\Gamma^{5678})
+a_4\Gamma^{3456}(1\pm\Gamma^{5678})
+a_6\Gamma^{5678}
+a_7\Gamma^{3457}(1\pm\Gamma^{5678})\,,
\label{eq:Fslash-28}
\end{equation}
with $a_1=f_1$, $a_2=f_2$, $a_4=f_{12}$, $a_7=f_{13}$ and $a_6=-(1\mp1)a_2a_4/a_1$ where the upper(lower) sign refers to the first(second) branch. Note that the last term anti-commutes with the $a_2$ and $a_4$-terms and commutes with the others. Diagonalizing the matrices as far as possible we can take
\begin{align}
\Gamma^{1234}\rightarrow\epsilon_1\mathbbm{1}\,,\qquad
\Gamma^{1256}\rightarrow\epsilon_2\sigma^3\,,\qquad
\Gamma^{3456}\rightarrow\epsilon_4\sigma^3\,,\qquad
\Gamma^{5678}\rightarrow\epsilon_6\mathbbm{1}\,,\qquad
\Gamma^{3457}\rightarrow\sigma^1\,,
\end{align}
when acting on $\xi$ with the sign factors satisfying $\epsilon_1\epsilon_2=-\epsilon_4$. Again we define $b_i=\epsilon_ia_i$.

Recalling the general form of the second supersymmetry condition in (\ref{eq:susy12}) given in (\ref{eq:susy2-general}) where $\slashed F=\slashed F_1+\slashed F_2$ and $[\slashed F_1,\Gamma_{\u b}]=0=\{\slashed F_2,\Gamma_{\u b}\}$ for a given $\u b$ and taking the difference of the equation for $\u b=5$ and $\u b=6$ we get the equation
\begin{equation}
(1\mp\epsilon_6)
\Big[
4b_1
-2b_6
+(1\pm1)b_2\sigma^3
\Big]\xi
=0\,.
\label{eq:28-5-6}
\end{equation}
From the equation for $f_1^2=b_1^2$ in terms of the others below (\ref{eq:F-28-2}) and (\ref{eq:F-28-3}) it is easy to verify that the determinant of the factor in square brackets cannot vanish implying that
\begin{equation}
\epsilon_6=\pm1\,,
\end{equation}
where the sign is correlated with the sign in (\ref{eq:Fslash-28}). Repeating the same calculation with $\u b=5$ and $\u b=7$ we find
\begin{equation}
(1\pm1)b_2\Big[2b_1-b_6+b_4\sigma^3+a_7\sigma^1\Big]\xi=0\,.
\label{eq:28-5-7}
%
%
%
%
\end{equation}
If we pick the upper sign, i.e. the second branch (\ref{eq:F-28-2}), we find, using the fact that in this case $b_6=0$ and $b_1^2=b_2^2+a_7^2+b_4^2$, that $b_2=f_2=0$ which again reduces us to background 29. We must therefore pick the lower sign so that we are dealing with the third branch (\ref{eq:F-28-3}) and we have the relations
\begin{equation}
b_1^2=b_2^2+b_4^2+2b_6^2+a_7^2\,,\qquad b_6=-a_6=-2\frac{b_2b_4}{b_1}\,.
\label{eq:constr-28}
\end{equation}

Taking $\u b=5$ we find that $\slashed F_1$ and $\slashed F_2$ commute and equation (\ref{eq:susy2-general}) factorizes giving
\begin{equation}
\slashed F_1(\slashed F_1-2b_1+2b_6)\xi=0\,,\qquad \slashed F_1=b_1+(b_2+b_4)\sigma^3+a_7\sigma^1\,.
\label{eq:28-56}
\end{equation}
The remaining equations come from taking $\u b=1$ and $\u b=3$ in the second equation in (\ref{eq:susy12}) and using again (\ref{eq:susy2-general}) they read
\begin{align}
\big(
b_4^2
+\tfrac14b_6^2
+a_7^2
-\tfrac12b_1b_6
-2b_2b_4
-b_6(b_2-b_4)\sigma^3
-b_1b_4\sigma^3
+(b_6-b_1)a_7\sigma^1
-b_2a_7\sigma^3\sigma^1
\big)\xi
=&\,0
\label{eq:28-12}
\\
\big(b_2^2
+\tfrac14b_6^2
-\tfrac12b_1b_6
-2b_2b_4
+b_6(b_2-b_4)\sigma^3
-b_1b_2\sigma^3
-b_6a_7\sigma^1
+b_2a_7\sigma^3\sigma^1
\big)\xi
=&\,0\,.
\label{eq:28-34}
\end{align}
Their sum gives
\begin{equation}
\big(
b_1^2
+b_2^2
+b_4^2
+\tfrac12b_6^2
+a_7^2
-b_1b_6
-4b_2b_4
-b_1\slashed F_1
\big)\xi=0\,.
\label{eq:sum12}
\end{equation}
Assuming that $\det\slashed F_1\neq0$ we have
\begin{equation}
(\slashed F_1-2b_1+2b_6)\xi=0
\end{equation}
and using this in the above equation gives
\begin{equation}
-b_1^2
+b_2^2
+b_4^2
+\tfrac12b_6^2
+a_7^2
+b_1b_6
-4b_2b_4
=0\,,
\end{equation}
which together with (\ref{eq:constr-28}) gives
\begin{equation}
b_6(2b_1-b_6)=0\,.
\end{equation}
The second factor cannot vanish so we must have $b_6=0$ but then $\det\slashed F_1=b_6(2b_6+b_1)=0$ in contradiction to our assumption. We conclude that $\det\slashed F_1=0$ so that 
\begin{equation}
b_6(2b_6+b_1)=0\,.
\end{equation}
If $\det(\slashed F_1-2b_1+2b_6)\neq0$ then $\slashed F_1\xi=0$ but in that case equation (\ref{eq:sum12}) is inconsistent with the above equation. We conclude that
\begin{equation}
\det(\slashed F_1-2b_1+2b_6)=3b_6(2b_6-b_1)=0
\end{equation}
and the only solution that solves also the previous equation is $0=b_6=-2b_2b_4/b_1$ so that $b_4=0$ ($b_2=0$ would again reduce us to the special case of background 29). The solution to (\ref{eq:constr-28}), (\ref{eq:28-56}), (\ref{eq:28-12}) and (\ref{eq:28-34}) becomes
\begin{equation}
b_1^2=b_2^2+a_7^2\,,\qquad(b_1-b_2\sigma^3-a_7\sigma^1)\xi\,.
\end{equation}
We find that $\slashed F$ takes the form (recall that we have $a_1=f_1$, $a_2=f_2$ and $a_7=f_{13}$ and the rest vanishing in (\ref{eq:F-28-3}))
\begin{equation}
\slashed F=a_1\Gamma^{1234}\big(2(1-2\mathcal P_1)\mathcal P_2+1\big)\,,
\end{equation}
where we have defined two spinor projection operators that commute and each project on a $16$-dimensional subspace
\begin{equation}
\mathcal P_1=\frac12\left(1+\frac{a_2}{a_1}\Gamma^{3456}+\frac{a_7}{a_1}\Gamma^{1257}\right)\,,\qquad(a_1^2=a_2^2+a_7^2)\,,\qquad
\mathcal P_2=\frac12(1-\Gamma^{5678})\,.
\label{eq:proj-28}
\end{equation}
Translating back to 32-component spinors the projection condition on the Killing spinor becomes
\begin{equation}
\xi=\mathcal P_1\mathcal P_2\xi\,,
\end{equation}
leaving eight non-zero components.

Using the fact that the non-zero components of $R_{\u a\u b}{}^{\u c\u d}$ are $R_{12}{}^{12}=-a_2^2$ and $R_{34}{}^{34}=-a_7^2$ it is not hard to verify that the remaining supersymmetry conditions (\ref{eq:susy3}) are also satisfied. This solution therefore preserves eight supersymmetries. Note that the special case $a_7=f_{13}=0$ corresponds to background $29$ which therefore preserves the same amount of supersymmetry.


\subsubsection*{60. $\fakebold{AdS_2\times S^3\times S^2\times T^4}$}
The flux takes the form (\ref{eq:F-60-1})--(\ref{eq:F-60-4})
\begin{equation}
\slashed F=
a_1\Gamma^{0123}
+a_2\Gamma^{0145}
+a_3\Gamma^{0167}
+a_4\Gamma^{2345}
+a_5\Gamma^{2367}
+a_6\Gamma^{4567}
+a_7\Gamma^{2346}(1\pm\Gamma^{4567})\,.
\end{equation}
If $a_7=f_{13}$ vanishes, which includes the first branch (\ref{eq:F-60-1}), $\slashed F$ is diagonalizable and comparing to (\ref{eq:F10}) the solution in (\ref{eq:n=8-AdS}) gives
\begin{equation}
b_1=b_2=b_3\,,
\end{equation}
and the rest vanishing, which reduces the background to 61. This solution turns out to be a special case of the more general solution with $a_7\neq0$ which will be found below.

It remains to analyze the branches (\ref{eq:F-60-2}), (\ref{eq:F-60-3}) and (\ref{eq:F-60-4}) with $a_7=f_{13}\neq0$. The flux takes the form
\begin{equation}
\slashed F=
a_1\Gamma^{0123}
+a_2\Gamma^{0145}(1-\Gamma^{4567})
+a_4\Gamma^{2345}(1\pm\Gamma^{4567})
+a_6\Gamma^{4567}
+a_7\Gamma^{2346}(1\pm\Gamma^{4567})\,,
\label{eq:Fslash-60}
\end{equation}
with $a_1=f_1$, $a_2=f_2$, $a_4=f_{12}$, $a_6=f_4$, $a_7=f_{13}$ and the condition $(1\pm1)a_1a_6=0$ whose solution distinguishes the first and second branch of solutions. Here the upper sign refers to the first two branches and the lower sign to the third branch. Since this is of the same form as (\ref{eq:Fslash-28}) above with only the indices on the gamma matrices shifted the supersymmetry equations coming from (\ref{eq:susy12}) take the same form as in (\ref{eq:28-5-6}), (\ref{eq:28-5-7}) and (\ref{eq:28-56})--(\ref{eq:28-34}) but with $b_{1,2}\rightarrow ib_{1,2}$ to account for the fact that $(\Gamma^0)^2=-1$. The first equation becomes
\begin{equation}
(1\mp\epsilon_6)
\Big[
4ib_1
-2b_6
+(1\pm1)ib_2\sigma^3
\Big]\xi
=0\,.
\label{eq:60-5-6}
\end{equation}
From the equation for $f_2^2=b_2^2$ in terms of the others below (\ref{eq:F-60-2}), (\ref{eq:F-60-3}) and (\ref{eq:F-60-4}) it is easy to verify that the determinant of the factor in square brackets can only vanish if we are in the branch (\ref{eq:F-60-3}) and $b_2^2=4b_1^2$. However in that case, assuming that the first factor is non-zero, i.e. $\epsilon_6=-1$, the second equation in (\ref{eq:susy12}) with $\u b=3$ implies $b_1=0$ which is inconsistent. Therefore we conclude again that the first factor must vanish and
\begin{equation}
\epsilon_6=\pm1\,,
\end{equation}
where the sign is correlated with the sign in (\ref{eq:Fslash-60}). Repeating the same calculation with $\u b=5$ and $\u b=7$ we find
\begin{equation}
(1\pm1)b_2\Big[2ib_1-b_6+b_4\sigma^3+a_7\sigma^1\Big]\xi=0\,.
\label{eq:60-5-7}
\end{equation}
Taking the upper sign, corresponding to the branches (\ref{eq:F-60-2}) and (\ref{eq:F-60-3}), we find that since $b_2$ cannot vanish the determinant of the factor in square brackets must vanish. This is only possible for the first branch but in that case the solution again becomes inconsistent with the second equation in (\ref{eq:susy12}) with $\u b=3$. We conclude that we must pick the lower sign corresponding to the branch (\ref{eq:F-60-4}) which has
\begin{equation}
b_2^2=b_1^2+b_4^2+2b_6^2+a_7^2\,,\qquad b_4=\frac{b_1b_6}{2b_2}\,.
\end{equation}

Replacing $b_{1,2}\rightarrow ib_{1,2}$ in (\ref{eq:28-56}) we find
\begin{equation}
\slashed F_1(\slashed F_1-2ib_1+2b_6)\xi=0\,,\qquad \slashed F_1=ib_1+(ib_2+b_4)\sigma^3+a_7\sigma^1\,.
\end{equation}
We must have either $\det\slashed F_1=0$ or $\det(\slashed F_1-2ib_1+2b_6)$ and in either case we find by looking at the imaginary terms that $b_2b_4=0$ and since $b_2$ cannot vanish that $b_4=0$. Looking at the real terms we then find that also $b_6=0$ and the remaining equations reduce to
\begin{equation}
\big(-ib_1+ib_2\sigma^3+a_7\sigma^1\big)\xi=0\,,\qquad b_2^2=b_1^2+a_7^2\,.
\end{equation}

We therefore find that $\slashed F$ takes essentially the same form as for background $28$ above (recall that $a_1=f_1$, $a_2=f_2$, $a_7=f_{13}$ and the rest vanishing in (\ref{eq:F-60-4}))
\begin{equation}
\slashed F=a_1\Gamma^{0123}\big(2(1-2\mathcal P_1)\mathcal P_2+1\big)\,,
\end{equation}
where the projectors are now
\begin{equation}
\mathcal P_1=\frac12\left(1+\frac{a_2}{a_1}\Gamma^{2345}-\frac{a_7}{a_1}\Gamma^{0146}\right)\,,\qquad(a_2^2=a_1^2+a_7^2)\,,\qquad
\mathcal P_2=\frac12(1-\Gamma^{4567})\,.
\label{eq:proj-60}
\end{equation}
The Killing spinor satisfies
\begin{equation}
\xi=\mathcal P_1\mathcal P_2\xi\,,
\end{equation}
leaving eight supersymmetries.

Using the fact that the non-zero components of $R_{\u a\u b}{}^{\u c\u d}$ are $R_{01}{}^{01}=a_2^2$ and $R_{23}{}^{23}=-a_7^2$ it is not hard to verify that the remaining supersymmetry conditions (\ref{eq:susy3}) are also satisfied. This solution therefore preserves eight supersymmetries. Note that the special case $a_7=f_{13}=0$ corresponds to background $61$ which therefore also preserves eight supersymmetries.


\subsubsection*{43. $\fakebold{AdS_2\times S^2\times S^2\times T^5}$}
The flux takes the form (\ref{eq:F-43-1})--(\ref{eq:F-43-6})
\begin{align}
\slashed F=&
a_1\Gamma^{0167}(1-\Gamma^{6789})
+a_2\Gamma^{2367}(1+\Gamma^{6789})
+a_3\Gamma^{2368}(1+\epsilon\Gamma^{6789})
+a_6\Gamma^{4567}(1+\Gamma^{6789})
\nonumber\\
&{}
+a_7\Gamma^{4568}(1\pm\Gamma^{6789})
+a_8\Gamma^{4569}(1\pm\Gamma^{6789})\,,
\end{align}
where the lower sign refers to the first two branches and the upper sign to the last four. The second and third branch have $\epsilon=-1$ and the rest have $\epsilon=+1$. Here $a_1=f_3\neq0$, $a_2=f_{12}$, $a_3=f_{13}$, $a_4=-f_{23}$, $a_5=f_{24}$, $a_6=g_{12}$, $a_7=g_{13}$, $a_8=g_{14}$, $a_9=g_{23}$, $a_{10}=g_{24}$. Effectively we can replace
\begin{align}
&\Gamma^{0167}\rightarrow i\sigma^1\,,\qquad
\Gamma^{2367}\rightarrow\epsilon_2\sigma^1\,,\qquad
\Gamma^{2368}\rightarrow\epsilon_3\sigma^2\,,\qquad
\Gamma^{4567}\rightarrow\epsilon_6\sigma^1\,,\qquad
\Gamma^{4568}\rightarrow\epsilon_7\sigma^2\,,
\nonumber\\
&\Gamma^{4569}\rightarrow\epsilon_8\sigma^3\,,\qquad
\Gamma^{6789}\rightarrow\epsilon_9\mathbbm{1}\,,\qquad
\end{align}
when acting on $\xi$ with suitable relations between the $\epsilon$'s. Taking $\u b=0$ in the second equation in (\ref{eq:susy12}) we get
\begin{equation}
\slashed F_1\big(\slashed F_1+ia_1(1-\epsilon_9)\sigma^1\big)\xi=0\,,\qquad
\slashed F_1=(1+\epsilon_9)(b_2+b_6)\sigma^1+[(1+\epsilon\epsilon_9)b_3+(1\pm\epsilon_9)b_7]\sigma^2+(1\pm\epsilon_9)b_8\sigma^3\,,
\end{equation}
while taking $\{\u a\u b\}=\{01\}$ in (\ref{eq:susy3}) we get
\begin{equation}
\big(
\slashed F_1^2
+8(1-\epsilon_9)a_1^2
+4i(1-\epsilon_9)a_1\sigma^1\slashed F_1
-36R_{01}{}^{01}
\big)\xi=0\,.
\end{equation}
Clearly we must take $\epsilon_9=-1$ since otherwise the curvature of $AdS_2$ would vanish. The above equations then become
\begin{equation}
\slashed F_1\big(\slashed F_1+2ia_1\sigma^1\big)\xi=0\,,\qquad
\big(8a_1^2+5ia_1\sigma^1\slashed F_1-18R_{01}{}^{01}\big)\xi=0\,,
\end{equation}
where $\slashed F_1=[(1-\epsilon)b_3+(1\mp1)b_7]\sigma^2+(1\mp1)b_8\sigma^3$. Multiplying the last equation by $\slashed F_1$ and using the first we find
\begin{equation}
\slashed F_1(a_1^2-R_{01}{}^{01})\xi=0\,,
\end{equation}
implying, using $\slashed F=\slashed F_1+2ia_1\sigma^1$, that
\begin{equation}
(\slashed F-2ia_1\sigma^1)\xi=0\,,\qquad R_{01}{}^{01}=\frac49a_1^2
\qquad\mbox{or}\qquad
\slashed F\xi=0\,,\qquad R_{01}{}^{01}=a_1^2\,.
\end{equation}
Only the latter curvature agrees with the Einstein equation. We conclude that $\slashed F\xi=0$ which implies
\begin{equation}
0=\det\slashed F=4a_1^2-[(1-\epsilon)b_3+(1\mp1)b_7]^2-2(1\mp1)b_8^2\,.
\end{equation}
This rules out the last three branches (\ref{eq:F-43-4})--(\ref{eq:F-43-6}) and reduces the first and third branch (\ref{eq:F-43-1}) and (\ref{eq:F-43-3}) to the third branch of background 45, (\ref{eq:F-45-3}), which will be a special case of the general solution found below. That leaves only the second branch (\ref{eq:F-43-2}) which satisfies the above condition. For this branch we have (recall that $a_1=f_3$, $a_3=f_{13}$ and $a_8=g_{14}$ and the rest vanishing)
\begin{equation}
\slashed F=4a_1\Gamma^{0167}(1-\mathcal P_1)\mathcal P_2
\,,\qquad
\xi=\mathcal P_1\mathcal P_2\xi\,,
\end{equation}
where
\begin{equation}
\mathcal P_1=\frac12\left(1-\frac{a_3}{a_1}\Gamma^{012378}-\frac{a_8}{a_1}\Gamma^{014579}\right)\,,\qquad
(a_1^2=a_3^2+a_8^2)\,,\qquad
\mathcal P_2=\frac12(1-\Gamma^{6789})\,.
\label{eq:proj-43}
\end{equation}
It is easy to check, using $R_{01}{}^{01}=a_1^2$, $R_{23}{}^{23}=-a_3^2$ and $R_{45}{}^{45}=-a_8^2$ that all the supersymmetry conditions are satisfied so that the background (\ref{eq:F-43-2}) with the above conditions preserves 8 supersymmetries. The special case $a_8=0$ corresponds to the third branch of background 45, (\ref{eq:F-45-3}), which therefore also preserves 8 supersymmetries.

\section*{{Acknowledgments}}
This work is supported by the STFC Consolidated grant ST/L00044X/1.

\newpage

\appendix

\section{Eleven-dimensional supergravity in superspace}\label{app:sugra}
In this appendix we will briefly recall the description of eleven-dimensional supergravity \cite{Cremmer:1978km} in superspace \cite{Cremmer:1980ru,Brink:1980az}. The basic superfields are the frame fields, or supervielbeins, $E^A$ for $A=\{a,\alpha\}$ with $a=0,\ldots,10$ and $\alpha=1,\ldots,32$ and the $SO(1,10)$ spin connection superfield $\Omega^{ab}$ which is anti-symmetric in the indices. These fields are functions of the superspace coordinates $z^M=\{x^m,\,\theta^\mu\}$.\footnote{Tangent space indices are taken from the beginning of the alphabet and coordinate indices from the middle of the alphabet.} The torsion and curvature superfields are defined in terms of these as
\begin{equation}
T^A=dE^A+E^B\wedge\Omega_B{}^A\,,\qquad R_A{}^B=d\Omega_A{}^B+\Omega_A{}^C\wedge\Omega_C{}^B\,,
\end{equation}
where we have defined $\Omega_\beta{}^a=0=\Omega_b{}^\alpha$ and $\Omega_\beta{}^\alpha=-\frac14\Omega^{ab}(\Gamma_{ab})^\alpha{}_\beta$.\footnote{The $32\times32$ gamma-matrices satisfy the basic Fierz identity
$$
\Gamma^a_{(\alpha\beta}(\Gamma_{ab})_{\gamma\delta)}=0
$$
and their anti-symmetric products satisfy the duality relations
$$
\Gamma^{a_1\cdots a_n}=\frac{(-1)^{[n/2]}}{(11-n)!}\varepsilon^{a_1\cdots a_na_{n+1}\cdots a_{11}}\Gamma_{a_{n+1}\cdots a_{11}}\,.
$$
} They satisfy the following Bianchi identities
\begin{equation}
dT^A+T^B\wedge\Omega_B{}^A=E^B\wedge R_B{}^A\,,\qquad 
dR_A{}^B+R_A{}^C\wedge\Omega_C{}^B-\Omega_A{}^C\wedge R_C{}^B=0\,.
\end{equation}
It is also convenient to introduce the four-form superfield $F=dC$ with $C$ the three-form superfield potential. Its Bianchi identity is
\begin{equation}
dF=0\,.
\end{equation}
In components the Bianchi identities read
\begin{align}
&\nabla_{[A}T_{BC]}{}^D+T_{[AB}{}^ET_{|E|C]}{}^D=R_{[ABC]}{}^D\,,\qquad
\nabla_{[A}R_{BC]D}{}^E+T_{[AB}{}^FR_{|F|C]D}{}^E=0\,,\qquad
\nonumber\\
&\nabla_{[A}F_{BCDE]}+2T_{[AB}{}^FF_{|F|CDE]}=0\,.
\end{align}

To obtain a sensible supergravity theory one must impose constraints on components of the torsion and curvature so that one is left with the right field content. It turns out that in eleven dimensions it is enough to impose the basic dimension zero constraint on the torsion $T_{\alpha\beta}{}^a\sim\Gamma^a_{\alpha\beta}$ \cite{Howe:1997he}. Everything else follows from the Bianchi identities.\footnote{This is not true, for example, in the case of type II supergravity in ten dimensions \cite{Wulff:2016tju}.} The constraints that follow from solving the Bianchi identities, organized according to the mass dimension of the fields, are
\begin{align}
\mbox{Dim 0: }&
T_{\alpha\beta}{}^a=-i(\Gamma^a)_{\alpha\beta}\,,\qquad
F_{\alpha\beta ab}=-i(\Gamma_{ab})_{\alpha\beta}\,,
\label{eq:dim0}
\\
\mbox{Dim 1: }&
T_{a\beta}{}^\gamma\equiv(G_a)^\gamma{}_\beta=\tfrac18(\slashed F\Gamma_a)^\gamma{}_\beta-\tfrac{1}{24}(\Gamma_a\slashed F)^\gamma{}_\beta\,,
\qquad(\slashed F\equiv\frac{1}{4!}F_{abcd}\Gamma^{abcd})
\\
&R_{\alpha\beta}{}^{ab}\equiv(G^{ab})_{\alpha\beta}=-2i(\Gamma^bG^a)_{(\alpha\beta)}
=
\tfrac{i}{4}(\Gamma^{[a}\slashed F\Gamma^{b]})_{\alpha\beta}
-\tfrac{i}{24}(\Gamma^{ab}\slashed F)_{\alpha\beta}
-\tfrac{i}{24}(\slashed F\Gamma^{ab})_{\alpha\beta}\,,
\nonumber\\
\mbox{Dim $\frac32$: }&
\Gamma^a\psi_{ab}=0\,,\qquad
R_{\alpha bcd}=\tfrac{i}{2}(\Gamma_b\psi_{cd})_\alpha-i(\Gamma_{[c}\psi_{d]b})_\alpha\,,\qquad(T_{ab}{}^\gamma\equiv\psi_{ab}^\gamma)\,,
\nonumber\\
&\nabla_\alpha F_{abcd}=6i(\Gamma_{[ab}\psi_{cd]})_\alpha\,,
\label{eq:dalphaF}
\\
\mbox{Dim 2: }&
\nabla_\alpha\psi_{ab}^\beta\equiv (U_{ab})^\beta{}_\alpha
=
2(G_{[a}G_{b]})^\beta{}_\alpha
-\tfrac14R_{ab}{}^{cd}(\Gamma_{cd})^\beta{}_\alpha
-2(\nabla_{[a}G_{b]})^\beta{}_\alpha\,,
\label{eq:Uab}
\\
\mbox{Dim $\frac52$: }&
\nabla_\alpha R_{ab}{}^{cd}=
-(G^{cd}\psi_{ab})_\alpha
-i(\Gamma_{[a}\nabla_{b]}\psi^{cd})_\alpha
-i(\psi^{cd}\Gamma_{[a}G_{b]})_\alpha
+2i(\psi_{[a}{}^{[c}\Gamma^{d]}G_{b]})_\alpha
\nonumber\\
&{}\qquad\qquad+2i(\Gamma^{[c}\nabla_{[a}\psi_{b]}{}^{d]})_\alpha\,,
\end{align}
while the remaining components, $T_{\alpha\beta}{}^\gamma$, $T_{ab}{}^c$ and $F_{\alpha bcd}$ vanish.

The supergravity equations of motion for the bosonic superfields follow from the fact that $\psi_{ab}$ is gamma-traceless, implying that $\Gamma^aU_{ab}=0$, and from the definition of $U_{ab}$ in (\ref{eq:Uab}) one finds
\begin{align}
&R_{ac}{}^{bc}=\frac{1}{144}\delta_a^bF_{cdef}F^{cdef}-\frac{1}{12}F_{acde}F^{bcde}\,,\qquad R_{[abc]d}=0\,,\nonumber\\
&\nabla_aF^{abcd}=-\frac{1}{2(4!)^2}\varepsilon^{bcdefghijkl}F_{efgh}F_{ijkl}\,,\qquad\nabla_{[a}F_{bcde]}=0\,,
\label{eq:eom}
\end{align}
or equivalently (\ref{eq:Einstein}).

\section{Symmetric space superisometry algebra}\label{app:iso}
Here we will derive the form of the superisometry algebra for a symmetric space background in terms of the geometry and flux. The derivation follows the same lines as in the $D=10$ type II case \cite{Wulff:2015mwa}.

A supergravity background has (super)isometries provided that there exist vector superfields
\begin{equation}
K_{(\epsilon)}=\epsilon^{\mathcal A}K_{\mathcal A}{}^M\partial_M\,,
\end{equation}
where $\epsilon^{\mathcal A}$ is a collection of constant infinitesimal parameters, such that the superspace Killing equation
\begin{equation}
\mathcal L_{K_{(\epsilon)}}E^A=E^Bl_B{}^A\,,
\label{eq:KE}
\end{equation}
 is satisfied, where $\mathcal L$ is the Lie derivative. This equation says that under an infinitesimal (super)isometry transformation the frame fields (supervielbeins) transform by a local Lorentz transformation $l_B{}^A$ whose non-zero components are $l_{ab}=-l_{ba}$ and $l_\beta{}^\alpha=-\frac14l_{ab}(\Gamma^{ab})^\alpha{}_\beta$. It is possible to pick the frame such that $l_{ab}=0$ but we will not do this here. Using the definition of the Lie derivative and of the torsion the superspace Killing equation takes the form
\begin{equation}
\nabla K_{(\epsilon)}^A+i_{K_{(\epsilon)}}T^A-E^B\big(l_B{}^A+i_{K_{(\epsilon)}}\Omega_B{}^A\big)=0\,.
\label{eq:KE2}
\end{equation}
Since we have introduced the four-form superfield strength $F$ we should also impose that it respects the isometries, i.e.
\begin{equation}
\mathcal L_KF=di_KF=0\qquad\mbox{or}\qquad K^E\nabla_E F_{ABCD}=-4(l_{[A}{}^E+i_K\Omega_{[A}{}^E)F_{|E|BCD]}\,.
\label{eq:F-iso}
\end{equation}
This condition is in fact implied by (\ref{eq:KE}) since the degrees of freedom of $F$ are contained in $E^A$.

Now we restrict to the case of symmetric spaces. In that case we have
\begin{equation}
\epsilon^{\mathcal A}=(\epsilon^a,\,\epsilon^{ab},\,\epsilon^{\hat\alpha})\,,
\end{equation}
corresponding to translations, rotations and supersymmetries. Note that $\epsilon^a$ exist for all values of $a=0,\ldots,10$ while $\epsilon^{ab}$ exist only for $a,b$ belonging to the same indecomposable factor in the geometry. The supersymmetry parameters $\epsilon^{\hat\alpha}$ exist for $\hat\alpha=1,\ldots,N$ corresponding to $N$ supersymmetries.\footnote{We have seen that for AdS backgrounds the possible values are $N=\{0,8,16,32\}$.} It is important to note that in addition the matrices defined by
\begin{equation}
K_a{}^m\,,\qquad K_{\hat\alpha}{}^{\hat\beta}=K_{\hat\alpha}{}^ME_M{}^{\hat\beta}\,,
\end{equation}
must have maximal rank, i.e. $11$ and $N$ respectively, by definition. Here we have split the Grassmann odd cotangent space as $E^\alpha=(E^{\hat\alpha},\,E^{\alpha'})$ where $E^{\hat\alpha}$ are the fermionic directions associated with supersymmetries. We can pick the Grassmann odd coordinates so that they also reflect this splitting, i.e. $\theta^\mu=\{\vartheta^{\hat\mu},\,\upsilon^{\mu'}\}$. This splitting will be such that $E^{\alpha'}|_{\upsilon=0}=0$. We will show below that the restriction of the full $(11|32)$-dimensional superspace to the $(11|N)$-dimensional subspace with $\upsilon=0$ gives a supercoset space and determine the corresponding superisometry algebra.

Consider the equation expressing the invariance of $F$ under an isometry transformation, (\ref{eq:F-iso}). We can pick a suitable frame such that the RHS vanishes and we have $K^E\nabla_E F_{ABCD}=0$. In the symmetric space case this gives the equations
\begin{equation}
K_a{}^F\nabla_F F_{bcde}=0\,,\qquad
K_{ab}{}^F\nabla_F F_{bcde}=0\,,\qquad
K_{\hat\alpha}{}^F\nabla_F F_{bcde}=0\,.
\end{equation}
The first and last equation give, using the invertibility of $K_a{}^b$ and $K_{\hat\alpha}{}^{\hat\beta}$ and restricting to the $\upsilon=0$ subspace with $\nabla_{\alpha'}F_{bcde}|_{\upsilon=0}=0$,
\begin{equation}
\nabla_AF_{bcde}|_{\upsilon=0}=0\,.
\end{equation}
The superspace constraint (\ref{eq:dalphaF}) then implies
\begin{equation}
\nabla_aF_{bcde}|_{\upsilon=0}=0\,,\qquad\psi_{ab}^\alpha|_{\upsilon=0}=0\,.
\end{equation}

In general one can determine the full superspace geometry by solving the equations (for more details see for example \cite{Wulff:2013kga})
\begin{equation}
\frac{d}{dt}E^A=i_\theta T^A\,,\qquad\frac{d}{dt}\Omega^{ab}=i_\theta R^{ab}\,,\qquad
\frac{d}{dt}F_{abcd}=\theta^\alpha\nabla_\alpha F_{abcd}\,,
\end{equation}
where one thinks of the superfields as functions of a parameter $t$ by rescaling the fermions as $\theta\rightarrow t\theta$. The boundary conditions for the solution are given by the bosonic geometry. In general these equations are very hard to solve (they were solved to order $\theta^5$ in \cite{Tsimpis:2004gq}) but they become very simple in our case when we restrict to the $\upsilon=0$ subspace. Using the superspace constraints (\ref{eq:dim0})--(\ref{eq:dalphaF}) we find (we drop the vertical bars in the following)
\begin{equation}
\frac{d}{dt}E^a=-iE\Gamma^a\vartheta\,,\qquad
\frac{d}{dt}E^{\hat\alpha}=E^a(G_a\vartheta)^{\hat\alpha}\,,\qquad
\frac{d}{dt}\Omega^{ab}=EG^{ab}\vartheta\,,\qquad
\frac{d}{dt}F_{abcd}=0\,.
\label{eq:supergeom}
\end{equation}
The equations are now linear and easy to solve. In fact these equations imply that the $(11|N)$-dimensional superspace obtained by setting $\upsilon=0$ is a supercoset space. To see this we define a Maurer-Cartan form valued in the superisometry algebra
\begin{equation}
J=\tfrac12\Omega^{ab}M_{ab}+E^aP_a+E^{\hat\alpha}Q_{\hat\alpha}\,.
\end{equation}
The equations for the supergeometry (\ref{eq:supergeom}) are easily shown to be equivalent to the Maurer-Cartan equation
\begin{equation}
dJ-J\wedge J=0\,,
\end{equation}
provided that the superisometry algebra takes the following form
\begin{align}
[P_a,P_b]=-\tfrac12R_{ab}{}^{cd}\,M_{cd}\,,\quad
[M_{ab},P_c]=2\eta_{c[a}P_{b]}\,,\quad
[M_{ab},M_{cd}]=2\eta_{c[a}M_{b]d}-2\eta_{d[a}M_{b]c}\,,
\nonumber\\
[P_a,Q_{\hat\alpha}]=(QG_a)_{\hat\alpha}\,,\quad
[M_{ab},Q_{\hat\alpha}]=-\tfrac12(Q\Gamma_{ab})_{\hat\alpha}\,,\quad
\{Q_{\hat\alpha},Q_{\hat\beta}\}=i\Gamma^a_{\hat\alpha\hat\beta}\,P_a-\tfrac12G^{ab}_{\hat\alpha\hat\beta}\,M_{ab}\,.
\label{eq:alg}
\end{align}
This is formally the same as in the type II case \cite{Wulff:2015mwa}. Note that the structure constants are determined in terms of the Riemann curvature tensor and the flux which are indeed constant for a symmetric space background and one can also verify that the Jacobi identity holds so that the superisometry algebra is indeed a Lie algebra. For the supersymmetric AdS solutions constructed here one obtains the algebras listed in table \ref{tab:susy-solutions}.

Let us finally note that using the results of this section it is easy to construct the supermembrane action restricted to the supercoset subspace. In the maximally supersymmetric cases this gives the full action which was already constructed long ago in \cite{deWit:1998yu}.

\section{No $AdS_2\times(S^2)^n\times T^{9-2n}$ solutions with flux on all flat directions}\label{app:ads2-s2}
Here we show that for the backgrounds $AdS_2\times S^2\times S^2\times S^2\times T^3$, $AdS_2\times S^2\times S^2\times T^5$ and $AdS_2\times S^2\times T^7$ it is not possible to have flux on all torus directions. This in turn means that these backgrounds must have $|F|^2=0$.

\noindent $\fakebold{AdS_2\times S^2\times S^2\times S^2\times T^3.}$ The flux takes the form 
\begin{align}
F=&
f_i\nu\wedge\sigma_i
+f_4\sigma_1\wedge\sigma_2
+f_5\sigma_1\wedge\sigma_3
+f_6\sigma_2\wedge\sigma_3
+f_7\nu\wedge\vartheta^{12}
+\sigma_1\wedge(f_8\vartheta^{12}+f_9\vartheta^{13})
\nonumber\\
&{}
+\sigma_2\wedge(f_{10}\vartheta^{12}+f_{11}\vartheta^{13}+f_{12}\vartheta^{23})
+\sigma_3\wedge(f_{13}\vartheta^{12}+f_{14}\vartheta^{13}+f_{15}\vartheta^{23})\,.
\end{align}
The condition $F\wedge F=0$ implies
\begin{align}
&f_1f_{12}=f_1f_{15}=0\,,\qquad
f_4f_3+f_5f_2+f_6f_1=0\,,\qquad
f_2f_8+f_7f_4+f_1f_{10}=0\,,\qquad
\nonumber\\
&f_2f_9+f_1f_{11}=0\,,\qquad
f_7f_5+f_3f_8+f_1f_{13}=0\,,\qquad
f_3f_9+f_1f_{14}=0\,,\qquad
\nonumber\\
&
f_7f_6+f_3f_{10}+f_2f_{13}=0\,,\qquad
f_3f_{11}+f_2f_{14}=0\,,\qquad
f_3f_{12}+f_2f_{15}=0\,,\qquad
\nonumber\\
&
f_4f_{13}+f_5f_{10}+f_6f_8=0\,,\qquad
f_4f_{14}+f_5f_{11}+f_6f_9=0\,,\qquad
f_4f_{15}+f_5f_{12}=0
\end{align}
and the Einstein equation in the flat directions imply
\begin{align}
&
f_1^2+f_2^2+f_3^2-f_4^2-f_5^2-f_6^2+3f_{12}^2+3f_{15}^2=0\,,\qquad
-f_7^2+f_8^2+f_{10}^2+f_{13}^2-f_{12}^2-f_{15}^2=0\,,
\nonumber\\
&f_9^2+f_{11}^2+f_{14}^2-f_{12}^2-f_{15}^2=0\,,\qquad
f_{11}f_{12}+f_{14}f_{15}=0\,,\qquad
f_{10}f_{12}+f_{13}f_{15}=0\,,\qquad
\nonumber\\
&f_8f_9+f_{10}f_{11}+f_{13}f_{14}=0\,.
\end{align}
If $f_1\neq0$ we find $f_{12}=f_{15}=0$ which implies that there cannot be flux on all torus directions. Therefore we must have $f_1=0$. Assuming $f_9\neq0$ we find $f_2=f_3=0$ and to have flux on all flat directions $f_7=0$. Since $f_7$ vanishes we can perform rotations to set also $f_8=f_{12}=0$ but it is then easy to see that $f_{15}$ must vanish and there cannot be flux on all torus directions. It remains to analyze the case when $f_9=0$ in which case we may also set $f_{12}=0$ by a rotation. Since we must have $f_{15}\neq0$ we find $f_2=f_4=f_{13}=f_{14}=0$ and $f_5f_{11}=f_3f_{11}=0$. Since $f_{11}$ cannot vanish we get $f_3=f_5=f_{10}=0$ but then we must have $f_6\neq0$ giving $f_7=f_8=0$ and there is again no solution.

\noindent $\fakebold{AdS_2\times S^2\times S^2\times T^5.}$ The flux takes the form
\begin{equation}
F=f_1\nu\wedge\sigma_1+f_2\nu\wedge\sigma_2+f_3\sigma_1\wedge\sigma_2+\nu\wedge\alpha+\sigma_1\wedge\beta+\sigma_2\wedge\gamma+\delta\,,
\end{equation}
with $\alpha,\beta,\gamma\in\Omega^2(T^5)$ and $\delta\in\Omega^4(T^5)$. The condition $F\wedge F=0$ forces 
\begin{equation}
f_1\gamma+f_2\beta+f_3\alpha=0\,,\qquad
f_1\delta=-\alpha\wedge\beta\,,\qquad
f_2\delta=-\alpha\wedge\gamma\,,\qquad
f_3\delta=-\beta\wedge\gamma\,.
\end{equation}
The trace of the Einstein equation in the flat directions gives
\begin{equation}
0=-|\alpha|^2+|\beta|^2+|\gamma|^2+2|\delta|^2-\tfrac56|F|^2=\tfrac16|F|^2+f_1^2+f_2^2-f_3^2+|\delta|^2\quad\Rightarrow\quad|F|^2=6(f_3^2-f_1^2-f_2^2-|\delta|^2)\,.
\end{equation}
Taking $\alpha=f_4\vartheta^{12}+f_5\vartheta^{34}$ the Einstein equation in the 5-direction gives
\begin{equation}
\tfrac12|i_5\beta|^2+\tfrac12|i_5\gamma|^2+\tfrac12|i_5\delta|^2+f_1^2+f_2^2+|\delta|^2-f_3^2=0\,,
\end{equation}
implying that to have flux on the 5-direction we need $f_3\neq0$. The $F\wedge F=0$ condition implies $\delta=-\frac{1}{f_3}\beta\wedge\gamma$, $f_1\gamma+f_2\beta+f_3\alpha=0$ and $f_2\beta^2=f_1\gamma^2=0$. Assuming $f_1\neq0$ find $\gamma=-(f_2/f_1)\beta-(f_3/f_1)\alpha$. We take, without loss of generality, $\beta=\frac12f_{ij}\vartheta^{ij}+f_6\vartheta^{15}+f_7\vartheta^{35}$ where $i,j=1,\ldots,4$.
The component Einstein equations $R_{15}=0$ and $R_{35}=0$ imply $f_{13}f_7=f_{13}f_6=0$ and since $f_6$ and $f_7$ cannot both vanish, since that would imply no flux in the 5-direction, we get $f_{13}=0$. Computing the combination of components of the Einstein equation in the flat directions $R_{11}-R_{22}+R_{33}-R_{44}$ gives
\begin{equation}
(-2f_{24}^2+f_6^2+f_7^2)(1+f_2^2/f_1^2)+f_4^2f_7^2/f_1^2+f_5^2f_6^2/f_1^2=0\,,
\end{equation}
showing that $f_{24}$ cannot vanish since that would imply $f_6=f_7=0$. The component Einstein equations $R_{12}=0$ and $R_{34}=0$ then imply $f_{14}=f_{23}=0$. And then $R_{13}=0$ implies $f_6f_7=0$ while computing the combination $R_{11}-R_{22}-R_{33}+R_{44}$ we find
\begin{equation}
f_6^2(1+f_2^2/f_1^2+f_5^2/f_1^2)-f_7^2(1+f_2^2/f_1^2+f_4^2/f_1^2)=0\,,
\end{equation}
whose only solution is $f_6=f_7=0$ in contradiction to the assumptions.

It remains to analyze the case $f_1=f_2=0$. $F\wedge F=0$ implies $\alpha=0$ and we may take $\beta=f_4\vartheta^{12}+f_5\vartheta^{34}$ and $\gamma=\frac12f_{ij}\vartheta^{ij}+f_6\vartheta^{15}+f_7\vartheta^{35}$ where $i,j=1,\ldots,4$. The analysis is essentially a special case of that carried out for $f_1\neq0$ and we conclude via the same steps that there is no solution with flux on all torus directions.

\noindent $\fakebold{AdS_2\times S^2\times T^7.}$ The flux takes the form
\begin{equation}
F=f_1\nu\wedge\sigma+\nu\wedge\alpha+\sigma\wedge\beta+\gamma\,,
\end{equation}
with $\alpha,\beta\in\Omega^2(T^7)$ and $\gamma\in\Omega^4(T^7)$. Taking the trace of the Einstein equation in the flat directions gives
\begin{equation}
0=-|\alpha|^2+|\beta|^2+2|\gamma|^2-\tfrac76|F|^2=f_1^2+|\gamma|^2-\tfrac16|F|^2\qquad\Rightarrow\qquad|F|^2=-6(f_1^2+|\gamma|^2)\,.
\end{equation}
Taking $\alpha=f_2\vartheta^{12}+f_3\vartheta^{34}+f_4\vartheta^{56}$ the Einstein equation in the 7-direction implies
\begin{equation}
0=\tfrac12|i_7\beta|^2+\tfrac12|i_7\gamma|^2-\tfrac16|F|^2=\tfrac12|i_7\beta|^2+\tfrac12|i_7\gamma|^2+f_1^2+|\gamma|^2\,,
\end{equation}
implying that there cannot be flux on the 7-direction.

\end{document}